\documentclass[aps,prl,twocolumn,groupedaddress,showpacs]{revtex4}

\usepackage{CJK}
\usepackage{amsmath}

\usepackage{graphicx}

\usepackage{dcolumn}

\usepackage{bm}
\usepackage[normalem]{ulem}
\usepackage{color}

\usepackage{epstopdf}

\newcommand{\ii}{{\rm i}}

\newcommand{\bq}{\mathbf{q}}

\newcommand{\br}{\mathbf{r}}
\newcommand{\bR}{\mathbf{R}}
\newcommand{\bff}{\mathbf{f}}

\newcommand{\bbr}{\mathbf{r}}
\newcommand{\bg}{\mathbf{g}}
\newcommand{\bh}{\mathbf{h}}

\newcommand{\bk}{\mathbf{k}}

\newcommand{\sep}{ \ \ \ , \ \ \ }

\newcommand{\beq}{\begin{equation}}
\newcommand{\eeq}{\end{equation}}
\newcommand{\beqn}{\begin{eqnarray}}
\newcommand{\eeqn}{\end{eqnarray}}
\newcommand{\pp}{\partial}
\newcommand{\dd}{{\rm d}}
\newcommand{\ee}{{\rm e}}

\newcommand{\eq}{Eq.\ }

\newcommand{\fig}{Fig.\ }

\newcommand{\la}{\langle}

\newcommand{\tbq}{\tilde{\bq}}
\newcommand{\tbk}{\tilde{\bk}}

\newcommand{\tbh}{\tilde{\bh}}
\newcommand{\ra}{\rangle}

\begin{document}

\title{Critical motility-induced phase separation belongs to the Ising universality class
}
\author{Benjamin Partridge}
\affiliation{Department of Bioengineering, Imperial College London, South Kensington Campus, London SW7 2AZ, U.K.}
\author{Chiu Fan Lee}
\email{c.lee@imperial.ac.uk}
\affiliation{Department of Bioengineering, Imperial College London, South Kensington Campus, London SW7 2AZ, U.K.}

\begin{abstract}
	A collection of self-propelled particles with volume exclusion interactions can exhibit the phenomenology of gas-liquid phase separation, known as motility-induced phase separation ({\bf MIPS}). The non-equilibrium nature of the system is fundamental to the phase transition, however, it is unclear whether MIPS at criticality contributes a novel universality class to non-equilibrium physics. We demonstrate here that this is not the case by showing that a generic critical MIPS belongs to the Ising universality class with conservative dynamics.
\end{abstract}
\pacs{}
\maketitle

Active matter is an extreme kind of non-equilibrium system in that detailed balance is broken at the microscopic scale \cite{marchetti_rmp13}. A typical active system can be a collection of particles that continuously exert mechanical forces on their surrounding environment, and systems of interacting active particles can display novel phenomena, ranging from the emergence of collective motion in two dimensions \cite{Vicsek1995, toner_prl95,toner_pre98} when the active particles are aligning,
to motility-induced phase separation ({\bf MIPS}) when the particles interact solely via volume exclusion interactions \cite{tailleur_prl08,fily_prl12, redner_prl13,	Cates2015,cates_annrev18}. However, even though active matter 
breaks detailed balance in a fundamental way, it remains unclear whether the hydrodynamic, universal behavior of active matter necessarily differs from that of equilibrium systems. 
Indeed, the ordered phase of a generic incompressible polar active fluid in 2D, and in 3D with an easy-plane, belong to the universality classes of equilibrium smectics in 2D \cite{chen_natcomm16} and the equilibrium sliding columnar phase \cite{chen_a18}, respectively. The investigation of universal behavior, besides being of central interest to physics, allows us to transfer knowledge of a well-known system to a different system of novel interest. Here, we do exactly that by demonstrating that the critical behaviour of MIPS belongs to the Ising universality class with conservative dynamics. We do so using three approaches: hydrodynamic argument, field-theoretic description of a microscopic model, and  simulation of a lattice model. 

\begin{figure}
	\begin{center}
		\includegraphics[scale=.38]{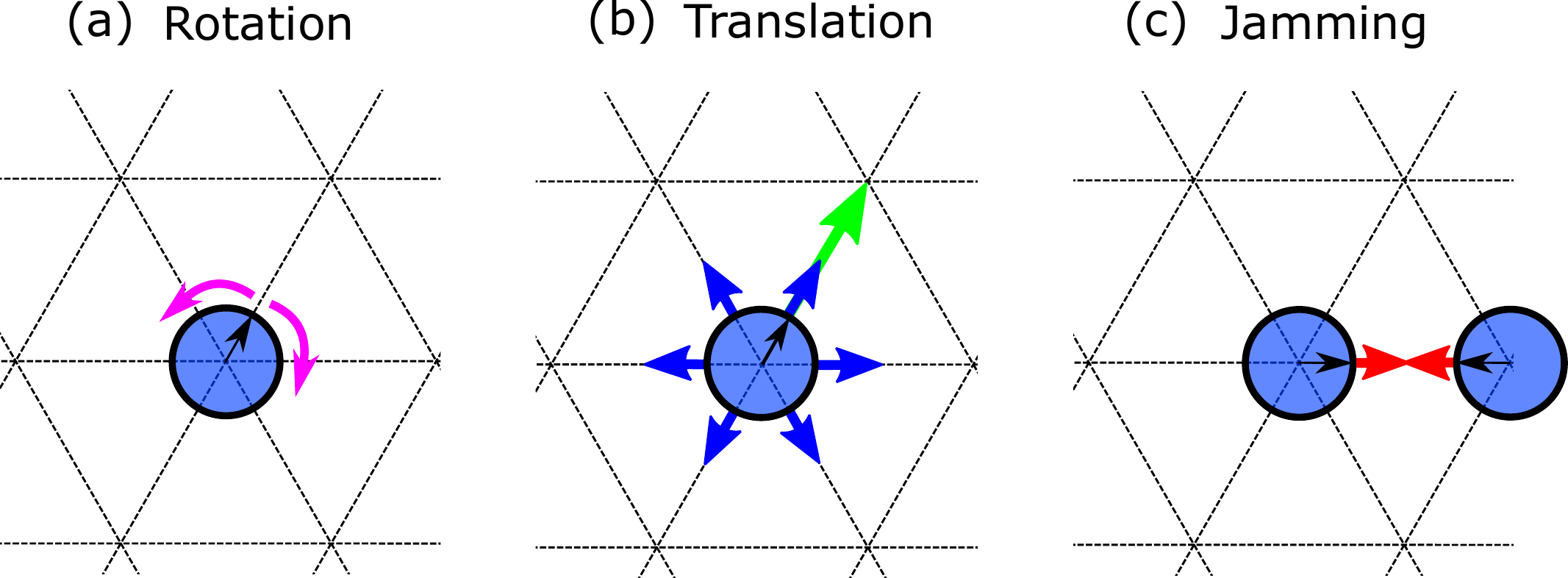}
	\end{center}
	\caption{
		{\it Active Brownian particles on a planar hexagonal lattice in our simulation.}
		Particle behavior is restricted to: (a) rotational diffusion, (b) ballistic motion (green arrow) and translation diffusion (blue arrows), and (c) disallowed translation due to attempting to move to an occupied lattice site. This minimal model generically exhibits MIPS.
	}
	\label{fig:cartoon}
\end{figure}

{\it Hydrodynamic argument.}
The generic system we are interested in consists of a collection of self-propelled particles in a frictional medium (i.e., no momentum conservation) with volume exclusion interactions (e.g., see \fig \ref{fig:cartoon}). As such it may be viewed as a typical compressible polar active fluid system \cite{Vicsek1995,toner_prl95,toner_pre98,gregoire_prl04}
with the alignment interactions switched off. In other words, a system undergoing MIPS constitutes a sub-class of an active fluid system that  is described by the Toner-Tu equations in the hydrodynamic limit \cite{toner_prl95,toner_pre98,toner_pre12}. Another way to view this is that since the symmetries underlying MIPS systems are the same as polar active fluids, the hydrodynamic equations, which are derived from symmetry consideration alone, must be the same. We therefore start with the Toner-Tu equations:
\begin{subequations}
	\label{eq1}
	\beqn
	\label{eq:rho}
	&& 	\pp_t \rho +\nabla \cdot \bg= 0
	\\
	\label{eq:bg}
	&& 	\pp_t \bg =-\zeta \nabla \rho -  \kappa \bg +\mu \nabla^2 \bg +\bff
	\eeqn
\end{subequations}
where only terms linear in the mass density field $\rho$ and the momentum density field $\bg$, together with their lowest order of spatial derivatives, are shown above since these terms suffice for our discussion. In (\ref{eq:bg}), $\bff$ is  a Gaussian noise with
spatio-temporal statistics:
\beq
\label{eq:noise}
\la \bff(t, \bbr) \ra=0 \ , \ \la \bff(t, \bbr) \bff(t', \bbr') \ra = 2D\delta(t-t')\delta(\bbr-\bbr')
\ ,
\eeq
where $D$ is the noise strength. 

Without the alignment interactions, collective motion is impossible. As a result, the momentum field has to go to zero in the hydrodynamic limit, implying that the coefficient $ \kappa $ 
has to be always positive. Therefore, the field $\bg$ is {\it not} a soft mode, namely, the momentum field is slaved to the density field. In the hydrodynamic limit, we can hence ignore the dynamical equation of $\bg$ and express $\bg$ as a function of $\rho$ and its derivatives:
\beq
\label{eq:exp_g}
\bg= -\nabla \Big[ (a_1 \phi +a_2 \phi^2 +a_3 \phi^3) - \left(\nabla^2 ( b \phi ) \right) +{\rm h.o.t} \Big]+\bff
\eeq
where $\phi(\br) = \rho(\br) -\rho_0$ for some constant $\rho_0$, 
and h.o.t.\ in (\ref{eq:exp_g}) refers to higher order terms in the expansion of $\bg$  in powers of $\nabla$ and $\phi$. Note that the negative sign in front of the square brackets is for  reasons of stability, and the noise term $\bff$ is as defined in (\ref{eq:noise}), albeit with the noise strength rescaled by $\kappa^{-2}$.

Substituting this form into (\ref{eq:rho}), we have
\beq
\label{eq:modelb}
\pp_t \phi = \nabla^2 \frac{\delta H}{\delta \phi} +{\rm h.o.t} \ +\nabla \cdot \bff
\eeq
where $H= \frac{a_1}{2} \phi^2+ \frac{a_2}{3} \phi^3+\frac{a_3}{4} \phi^4+\frac{b}{2} \left(\nabla \phi \right)^2$
is the familiar Landau-Ginzburg Hamiltonian while h.o.t.\ in (\ref{eq:modelb}) again refers to higher order terms omitted, which include non-equilibrium terms such as  $\nabla^4 \phi^2$  \cite{wittkowski_natcomm14} and $\nabla \cdot [\nabla \phi (\nabla^2 \phi)]$
\cite{Tjhung2018}. 
We note that although the resulting EOM is similar to the Active Model B introduced in \cite{wittkowski_natcomm14}, our approach is completely different -- 
	our EOM arises from premises based explicitly on symmetry consideration alone. Practically, our method  inevitably leads to the presence of the  $\nabla^2 \phi^2$ term in the EOM due to absence of Ising symmetry  (\(\phi \mapsto -\phi\)). Such a term is absent in the Active Model B.

\begin{figure}
	\begin{center}
		\includegraphics[scale=.54]{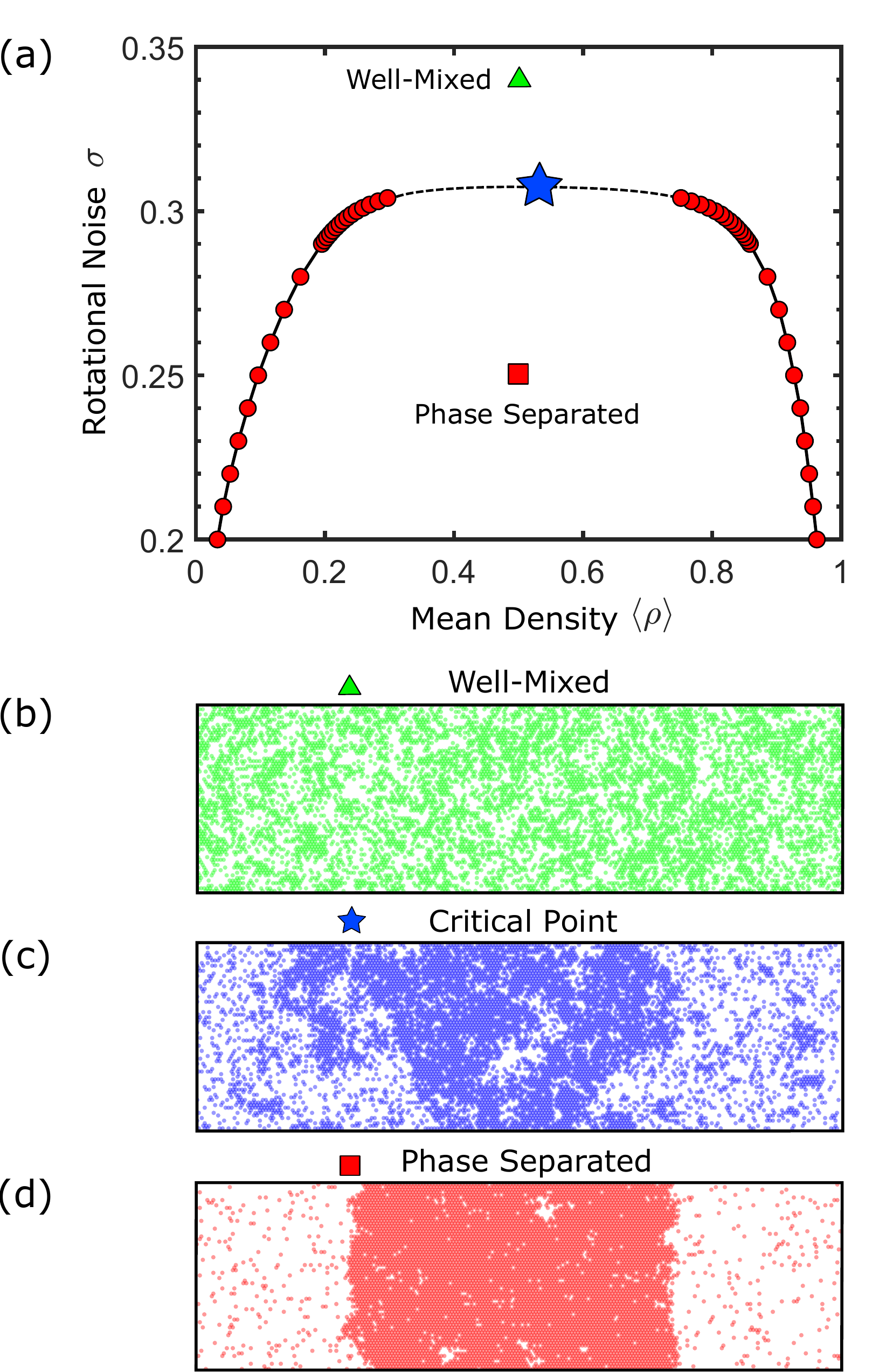}
	\end{center}
	\caption{{\it Phase behavior.} a) The phase diagram resulting from Monte Carlo simulations of the lattice model (\fig \ref{fig:cartoon}) shows the well-mixed region at high rotational noise $(\sigma)$ and the phase separated region at low noise. Snapshots of the system configuration at the well-mixed region, critical point (estimated  to be at $ \rho_c 
		\approx 0.522, \sigma \approx 0.305$ \cite{SI}), and phase separated region are shown in (b), (c), and (d), respectively. The  system shown has $72 \times 216$ sites. } 
	\label{fig:phases}
\end{figure}

\begin{figure*}
	\begin{center}
		\includegraphics[scale=.405]{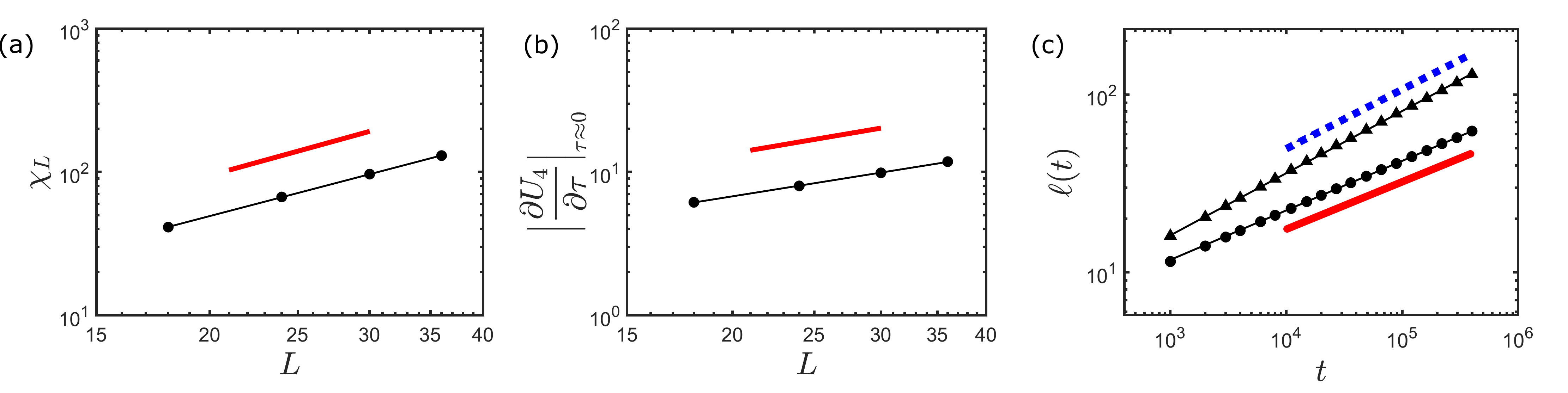}
	\end{center}
	\caption{
		{\it Static and dynamic exponents estimation from our lattice model simulations.}
		a) Susceptibility \(\chi_L\) at criticality as a function of system size \(L\); b) The Binder cumulant at criticality \((\partial_{\tau} U_4)_{\tau \simeq 0}\), where \(\tau\) is the dimensionless distance to the critical noise, as a function of $L$; c)  The average coarsening-length scale \(\ell(t)\) at criticality (circles) and deep within the phase-separated regime (triangles), as determined by the first intercept of the pair correlation function with the $x$-axis, vs.~time $t$ as measured by the number of particle sweeps. 
		The red lines show the exact results from the 2D Ising model with conservative dynamics: a) \(\gamma/\nu = 7/4\), b) $\nu = 1$ and c) $z = 15/4$. While the dashed blue line in c) shows the Lifshitz-Slyozov \(z = 3\) scaling. In each plot error bars are smaller than the size of the data points. See \cite{SI} for details on simulation procedures and error estimation.	
	}
	\label{fig:results}
\end{figure*}

Now, the phenomenology of MIPS indicates that the system can be placed at the critical point by tuning two model parameters, e.g., by tuning the density and the noise strength (\fig \ref{fig:phases}). Given this constraint,  the only possibility to achieve criticality (i.e., having a divergent correlation length in the system)  corresponds to tuning $a_1$ and $a_2$ to zero in $H$. Around this critical point, standard renormalization group method demonstrate that all higher order terms in (\ref{eq:modelb}) are irrelevant \cite{Cardy1996,Kardar2007}. In particular, the dynamical equation (\ref{eq:modelb}) is exactly the dynamics of the Ising model with conservative dynamics (model B) \cite{hohenberg_rmp77}.
The scaling behavior of MIPS at criticality  is thus characterized by three exponents: two static and one dynamic. 
Our hydrodynamic argument applies in {\it any} spatial dimension. To verify this conclusion, we will from now on focus on MIPS in 2D, and first look 
at a field-theoretic formulation of a specific lattice model in 2D to see how the system can be fine tuned to exhibit critical behavior. We will then 
demonstrate with simulation results that critical exponents of MIPS show good agreement with our prediction (\fig \ref{fig:results}). 
We note that in an interesting development, independently and concurrently to our work, it  has also been concluded \cite{caballero_2018} that Ising behavior is generically possible. However, the authors also speculate that different, non-equilibrium strong coupling behavior is possible, based on generalizing the perturbative RG analysis beyond the controlled regime. We do not see evidence of such a regime in our simulation results.


{\it Field-theoretic description of a microscopic model.} 
There are some notable attempts to represent MIPS on lattice \cite{thompson_2011,soto_2014,kourbanehoussene_2018,whitelam_2018}, which is especially useful in the study of the collective dynamics and the emergence of pattern formation in bacteria \cite{thompson_2011,soto_2014}. Here, we consider  an active particle model on a 2D hexagonal lattice similar to the one recently introduced in \cite{lee_njp13}, except here the occupancy of the lattice site is bounded by a constant $M$. Specifically, we consider a collection of six distinct types of active particles, each type has a specific orientation $\theta_i$ and will only jump to the neighboring site along the direction $\theta_i$,  with a certain rate that depends on the occupancy of the target site. 
In addition, the type of a particle will convert, with rate $\sigma$, to a different type in the neighboring orientations, which corresponds to the rotational noise of the particle. Enumerating the  lattice by the set of vectors $\bR$, and the type (orientation) of the active particle by $i$, we now denote the number of particles of type $i$ on the lattice site $\bR$ by $A_{\bR}^{\theta_i}$.

Using the field-theoretic formalism developed in \cite{lefevre_jstatmech07}, the action that describes the model is
\begin{subequations}
	\beqn
	&&S =  \int \dd t  \sum_\bR \Bigg\{ \sum_i \Bigg[ \hat{A}_{\bR}^{\theta_i} \pp_t A_{\bR}^{\theta_i}
	\\
	\label{eq:jumping}
	&&- 
	\gamma \left(M -N_{\bR + {\bf e}_{\theta_i}}\right) A_{\bR}^{\theta_i}
	\left(\ee^{\hat{A}_{\bR+ {\bf e}_{\theta_i}}^{\theta_i}
		-\hat{A}_\bR^{\theta_i}}-1\right)
	\Bigg]
	\\
	\label{eq:rotation}
	&&
	-\sum_{\la i,j \ra} \sigma A_\bR^{\theta_i} \left(\ee^{\hat{A}_\bR^{\theta_j} -\hat{A}_\bR^{\theta_i}}-1\right)
	\Bigg\}
	\ ,
	\eeqn
\end{subequations}
where ${\bf e}_\theta \equiv d(\cos \theta \hat{\bf x} + \sin \theta \hat{\bf y}$) is a vector pointing along the direction $\theta$ with its norm being the lattice spacing $d$, $\hat{A}_\bR^{\theta_i}$ are the conjugate fields of ${A}_\bR^{\theta_i}$,
and $N_{\bR }= \sum_j A_{\bR }^{\theta_j}$ is the total number of particles on site $\bR$. Specifically, \eq (\ref{eq:jumping}) corresponds to the jumping event with rate $\gamma (M -N_{\bR + {\bf e}_{\theta_i}})$, which means that the rate of jumping into a lattice decreases with the occupancy of that target lattice and becomes zero if the lattice site has already $M$ particles. This models the  volume exclusion interactions between the particles. \eq (\ref{eq:rotation}) corresponds to the interconversion between the particle types and $\la i,j \ra$ denotes the pairs of orientations that are nearest neighbors to each other in angular space.

By first taking the discrete spacing of the lattice to zero ($d \rightarrow 0$), and then the angular space to the continuum limit, we argue in \cite{SI} that the action can be approximated as
\beqn
\nonumber
S&=&
\int \dd t \dd^2 r \dd \theta  \Bigg\{ \hat{\psi}_{\theta} \Big[ \pp_t \psi_{\theta}+
\hat{\bf e}_{\theta} \cdot \nabla_\br \left[\gamma \left(\rho_M -\rho\right)  \psi_{\theta} \right]
\\
&&
- \sigma \pp_\theta^2 \psi_{\theta}\Big]  -\sigma \psi_\theta \left(\pp_\theta \hat{\psi}_{\theta}\right)^2
\Bigg\}
\ ,
\eeqn
where 
$\hat{\bf e}_\theta$ is now a normalized unit vector, $\psi_\theta(\br) \propto A^\theta_\bR/d^2$ is the density of particles with orientation $\theta$ at $\br =\bR$, $\hat{\psi}_\theta(\br) = \hat{A}^\theta_\bR$, $\rho(\br) = (2\pi)^{-1}\int \dd \theta  \psi_\theta (\br)$ is the particle density at position $\br$, and  $\rho_M \propto M/d^2$ is the maximal density allowed.

Since the action is now quadratic in $\hat{\psi}_\theta$, we can re-write the dynamics of this field-theoretic model as a Langevin equation:
\beq
\label{eq:psi}
\pp_t \psi_\theta+\hat{\bf e}_\theta \cdot \nabla_\bbr\left[\gamma(\rho_M-\rho) \psi_\theta \right]
=\sigma \pp_\theta^2 \psi_\theta +\xi_\theta 
\eeq
where 
$\xi_\theta$ are noise terms with statistics:
\beqn
\la \xi_\theta(\bbr, t)  \ra&=&0
\\
\nonumber
\la \xi_\theta(\bbr, t) \xi_{\theta'}(\bbr',t') \ra &=& 2 \sigma \pp_\theta^2 \left[ \psi_\theta
\delta(\theta-\theta') 
\delta(t-t') \delta^2(\bbr-\bbr') \right]
\ .
\eeqn

The set of EOM \eq (\ref{eq:psi}) constitutes an infinite number of field equations (one for each $\theta$). To reduce these into the hydrodynamic equations of the form (\ref{eq1}), we consider the Fourier expansion of $\psi$ with respect to $\theta$ \cite{lee_pre_CM10}: 
\beq
\psi_\theta(\bbr)= \alpha_0(\bbr)+2\sum_{n\geq 1} \left[ \alpha_n(\bbr) \cos (n \theta) +\beta_n (\bbr) \sin(n\theta)\right]
\ .
\eeq
In particular, $\alpha_0= \rho$,
$\alpha_n =\alpha_{-n}$ and $\beta_n =-\beta_{-n}$. In \cite{SI}, we show that in the Fourier transformed space, all modes are massive except for $\rho$, which is consistent with our previous hydrodynamic argument. Here, we aim to demonstrate  how the coefficients in the hydrodynamic equations can in principle be fine tuned, we will thus simplify the EOM by  setting $\beta_n$ to zero for all $n$ (hence spatial variation is only possible along the $x$-axis), and $\alpha_n$ to zero for all $n>1$. Note that this kind reduction has been shown to be useful in the study of polar active fluids at the onset of collective motion \cite{bertin_pre06,Peruani2008,bertin_jpa09,lee_pre_CM10}. 

Going through this reduction procedure \cite{SI}, we arrive at 
\begin{subequations}
	\beqn
	\pp_t \rho +  \frac{\pi \gamma}{2} \pp_x \left[(\rho_M-\rho) \alpha_1\right] &=& 0
	\\
	\pp_t \alpha_1 +  \frac{\pi \gamma}{2} \pp_x \left[(\rho_M-\rho)\rho\right]  &=& -\sigma \alpha_1 
	\ ,
	\eeqn
\end{subequations}
where we have also ingored the fluctuating term in $\alpha_1$, which we will discuss later.

Solving for $\alpha_1$ by setting the temporal derivative of $\alpha_1$ to zero (since it is a fast mode), we have
\beq
\label{eq:newrho}
\pp_t \rho +  \pp_x \left\{-\frac{\pi^2 \gamma^2}{4 \sigma}(\rho_M-\rho)\pp_x \Big[ (\rho_M-\rho)\rho\Big] \right\} = 0
\ .
\eeq
The expression inside the curly brackets corresponds to the $x$-component of $\bg$ in (\ref{eq:bg}). 
	Note that at this level of truncation, the above model equation is similar to other models \cite{solon_pre18, Tjhung2018}. However, we will show in	 \cite{SI} how our field-theoretic model leads to other non-equilibrium terms (e.g., $\nabla^4 \phi^2$ and $\nabla \cdot [(\nabla \phi) (\nabla^2 \phi)]$ in (\ref{eq:modelb})) when higher order modes are incorporated.

Expressing $\rho$ as $\rho_0+\phi$ for some constant $\rho_0$ in (\ref{eq:newrho}), we find 
\begin{subequations}
	\label{eq:as}
	\beqn
	a_1 &=&\frac{\pi^2 \gamma^2}{4\sigma} (\rho_M-\rho_0)(\rho_M-2\rho_0)
	\\
	a_2 &=&\frac{\pi^2 \gamma^2}{4\sigma} \frac{4\rho_0-3\rho_M}{2}
	\ .
	\eeqn
\end{subequations}
As aforementioned, the system exhibits Ising critical behaviour when both $a_1$ and $a_2$ are zero. By tuning $\rho_0$, we can either set  $a_1$ to zero (when $\rho_0 = \rho_M/2$), or $a_2$ to zero (when $\rho_0 = 3\rho_M/4$), but seemingly not both. However, we have not yet incorporated the noise term into the analysis. Indeed,  by analyzing the hydrodynamic equation
(\ref{eq:modelb}) using diagrammatic methods around the critical point, one finds that fluctuation-induced renormalization of the coefficients generically  increases $a_1$ while decreases $a_2$ \cite{SI}. In other words, the fluctuation strength can in principle be fined tuned so that both $a_1$ and $a_2$ are zero. In particular, for this microscopic model, the critical density $\rho_c$ is bounded below by $\rho_M/2$ and above by $3\rho_M/4$. We will see that this bound is also satisfied by our simulated system (\fig \ref{fig:phases}a), which we will turn to now.

{\it Simulation of a lattice model.}
Going beyond our analytical arguments, we will now present simulation results in support of our conclusion. 
We employ a similar  microscopic model as in our field-theoretic formulation (with maximal occupancy $M=1$), except that we allow the particles to diffuse, with a low probability, in addition to the active movement. This is done to improve computational efficiency. Data analysis is adapted from  \cite{siebert_a17}. Note that based on simulation results in continuum space for the static critical exponents, the authors in \cite{siebert_a17} arrived at a different conclusion from us. We speculate that the discrepancy arises because their results are not yet in the scaling regime, potentially due to the limited sizes used in the study. Here, by focusing on a lattice model, we can perform simulations on larger systems and achieve better statistics, enabling us to find good agreement between our analytical predictions and the simulation results for both the static exponents as well as the dynamic exponent (\fig \ref{fig:results}).

In our system,  \(N\) polar particles move on an elongated hexagonal lattice of size \(2L \times 6L\) lattice sites subject to periodic boundary conditions. 
The system evolves via
an iterative Monte-Carlo style update scheme in which particles are selected at random and we measure time $t$ in particle sweeps. Specifically, at each time step two stochastic processes per particle are attempted: 1) to implement rotational fluctuations of a particle, a Gaussian random variable
with standard deviation \( \sigma \) is drawn and rounded to the nearest integer $n$, the particle's direction is then rotated by $n\times 60^\circ$; 2) to implement translation, the particle will attempt to move in a direction  prescribed by its orientation with probability $24/30$ (active motion), and in a randomly chosen direction otherwise (diffusive motion). Steric interactions are
implemented by disallowing any movement into an occupied site.

Using the sampling method described in \cite{SI} we construct a time-series of density samples representative of the co-existing liquid and vapour phases. The resulting phase diagram is shown in \fig \ref{fig:phases}.
In particular we are interested in the fourth-order Binder cumulant \(U_4\), whose invariance with respect to system size \(L\) at criticality we use to locate the asymptotic critical noise strength.

Based on our previous analytical arguments, 
three independent critical exponents will characterise fully the universal behaviour of critical MIPS. Focusing first on the static critical exponents, 
we use the standard finite-size scaling relations \cite{landau_b2014}:
$\chi_L \sim L^{\gamma/\nu}$ and $|\partial U_4/\pp \tau | \sim L^{1/\nu}$, 
where \(\tau\) is the dimensionless distance to the critical noise and \(\chi_L\) is the finite-size sub-box susceptibility:
\begin{equation}
\chi_L = \frac{\langle N^2 \rangle_L - {\langle N \rangle_L}^2}{\langle N \rangle_L} \ .
\end{equation} 
In the above, \(\langle . \rangle_L\) denotes an average taken over a finite system size \(L\). 
The results of this analysis are shown in Figs \ref{fig:results}a \& b, which show good agreement with the  analytical results for the 2D Ising universality class (red lines).

To estimate the dynamic exponent $z$ for critical MIPS,
we adapt a method presented in \cite{alexander_prb94}:
we seed 
300 simulation runs from a completely disordered initial state and calculate the characteristic coarsening length \(\ell(t)\) of the system for the first 10,000 particle sweeps. We define \(\ell (t)\) as the length-scale at which the correlation function first becomes negative. Again, we see a good agreement between the data and the Ising result (\fig \ref{fig:results}c).
To further ascertain the validity of our simulation method, we repeat this procedure deep within the phase separated regime, where due to the emergence of the Gibbs-Thomson relation at the interface \cite{lee_softmatter17,solon_njp18}, we expect that the coarsening dynamics of MIPS  at the late stage follows the equilibrium Lifshitz-Slyozov scaling law ($z =3$), which is indeed the case (\fig \ref{fig:results}c).

{\it Conclusion.} We have demonstrated that the critical behaviour of MIPS does not lead to a novel universality class, rather, it  belongs generically to the equilibrium Ising universality class with conservative dynamics.
	Our hydrodynamic approach is based solely upon consideration of symmetry and conservation law. Therefore, our conclusion applies to all models consistent with these premises. In particular, since neither the mechanism of self-propulsion nor the particularity of the noise can effect the symmetry of the system, a broad class of dry active matter models displaying critical MIPS will belong to the Ising universality class. 	 
We also note that novel critical  behaviour in active matter is indeed possible \cite{chen_njp15} and it remains an interesting question to see what  universality classes unique to active matter await discovery.

\vspace{.05in}
\begin{acknowledgments}
	We thank John Toner (Oregon) for stimulating discussion, and acknowledge the High Throughput Computing service provided by Imperial College Research Computing Service,  DOI: 10.14469/hpc/2232.
\end{acknowledgments}


\onecolumngrid

\newpage

\begin{center}
	
	\textbf{\large Supplemental Materials:\\ Critical motility-induced phase separation belongs to the Ising universality class
	}
	\vspace{.1in}
	
	Benjamin Partridge \& 
	Chiu Fan Lee*\\
	{\it Department of Bioengineering, Imperial College London, South Kensington Campus, London SW7 2AZ, U.K.\\
		* Electronic address: c.lee@imperial.ac.uk}

\end{center}

\section{Field-theoretic description of a microscopic model} 
We start with the action (\eq (6)) from the Main Text ({\bf MT}):
\begin{subequations}
	\beqn
	&&S =  \int \dd t  \sum_\bR \Bigg\{ \sum_i \Bigg[ { \hat{A}_{\bR}^{\theta_i} \pp_t A_{\bR}^{\theta_i}}
	- 
	\gamma \left(M -N_{\bR + {\bf e}_{\theta_i}}\right)  { A_{\bR}^{\theta_i}}
	\left(\ee^{\hat{A}_{\bR+ {\bf e}_{\theta_i}}^{\theta_i}
		-\hat{A}_\bR^{\theta_i}}-1\right)
	\Bigg]
	\\
	&&
	-\sum_{\la i,j \ra} \sigma A_\bR^{\theta_i} \left(\ee^{\hat{A}_\bR^{\theta_j} -\hat{A}_\bR^{\theta_i}}-1\right)
	\Bigg\}
	\ ,
	\eeqn
\end{subequations}
where ${\bf e}_\theta \equiv d(\cos \theta \hat{\bf x} + \sin \theta \hat{\bf y}$) is a vector pointing along the direction $\theta$ with its norm being the lattice spacing $d$, $\hat{A}_\bR^{\theta_i}$ are the conjugate fields of ${A}_\bR^{\theta_i}$,
and $N_{\bR }= \sum_j A_{\bR }^{\theta_j}$ is the total number of particles on site $\bR$. 

Following \cite{Alefevre_jstatmech07}, we perform the following expansion:
\beq
\exp \left[{\hat{A}_{\bR + \hat{\bf e}_{\theta_j}}^{\theta_j} -\hat{A}_\bR^{\theta_i}}\right]-1 =d \hat{\bf e}_{\theta_j} \cdot \nabla \hat{A}_\bR^{\theta_i} + o (d)
\ ,
\eeq
where we have taken the continuum limit by taking the lattice spacing $d$ to zero, with $\br = \bR$,  and replaced $(\sqrt{3}/2)d^2\sum_\br$ by $\int \dd^2 r$ (the prefactor comes from the fact that we are considering a hexagonal lattice), $(2/\sqrt{3}) A_\bR^{\theta_i}/d^2$ by $\psi_{\theta_i}(\br)$, $\hat{A}_\bR^{\theta_i}$ by $\hat{\psi}_{\theta_i}(\br)$, $(2/\sqrt{3}) N_{\br + \hat{\bf e}_{\theta_j}}/d^2$ by $\rho(\br)$, and $(2/\sqrt{3}) M/d^2$ by $\rho_M$, which corresponds to the maximum density allowed. In addition, $\gamma$ is now rescaled by $d^2$.

With the truncated expansion, the action is now
\beq
\label{Aeq:action2}
S =  \int \dd t \dd^2 r \Bigg\{ \sum_i \Bigg[ \hat{\psi}_{\theta_i} \pp_t \psi_{\theta_i}
- 
\gamma \left(\rho_M -\rho\right)  \psi_{\theta_i}  \hat{\bf e}_{\theta_j} \cdot \nabla \hat{\psi}_{\theta_i}
\Bigg]
-\sigma\sum_{\la i,j \ra}  \psi_{\theta_i} \left(\ee^{\hat{\psi}_{\theta_j} -\hat{\psi}_{\theta_i}}-1\right)
\Bigg\}
\ ,
\eeq
where we have taken the continuum limit by taking the lattice spacing $d$ to zero, with $\br = \bR$,  and replaced $d^2\sum_\br$ by $\int \dd^2 r$, $A_\bR^{\theta_i}/d^2$ by $\psi_{\theta_i}(\br)$, $\hat{A}_\bR^{\theta_i}$ by $\hat{\psi}_{\theta_i}(\br)$, $N_{\br + \hat{\bf e}_{\theta_j}}/d^2$ by $\rho(\br)$, and $M/d^2$ by $\rho_M$, which corresponds to the maximum density allowed. In addition, $\gamma$ is now rescaled by $d^2$.

By taking this continuum limit, the original spatial lattice structure is lost and we will now assume that we can take the increment in the angular space to the infinitesimal limit, i.e., $|\theta_{i+1} - \theta_i| \rightarrow 0$. The last term in (\ref{Aeq:action2}) then becomes \cite{Alefevre_jstatmech07}:
\beq
\sigma\sum_{\la i,j \ra}  \psi_{\theta_i} \left(\ee^{\hat{\psi}_{\theta_j} -\hat{\psi}_{\theta_i}}-1\right) \mapsto \sigma {\psi}_{\theta} \left[ \pp_\theta^2 \hat{\psi}_{\theta}  + \left(\pp_\theta \hat{\psi}_{\theta}\right)^2 \right]
\ .
\eeq
Substituting the above into (\ref{Aeq:action2}), the action becomes
\beq
S=
\int \dd t \dd^2 r \dd \theta  \Bigg\{ \hat{\psi}_{\theta} \Big[ \pp_t \psi_{\theta}+
\hat{\bf e}_{\theta} \cdot \nabla_\br \left[\gamma \left(\rho_M -\rho\right)  \psi_{\theta} \right]
- \sigma \pp_\theta^2 \psi_{\theta}\Big]  -\sigma \psi_\theta \left(\pp_\theta \hat{\psi}_{\theta}\right)^2
\Bigg\}
\ .
\eeq


\section{Mode reduction from the field-theoretic equations of $\psi_\theta$}

We start with the EOM of $\psi_\theta(\br)$:
\beq
=\sigma \pp_\theta^2 \psi_\theta +\xi_\theta 
\eeq
where $\hat{\bf e}_\theta \equiv \cos \theta \hat{\bf x} + \sin \theta \hat{\bf y}$ and $\xi_\theta$ are noise terms with statistics:
\beqn
\la \xi_\theta(\bbr, t)  \ra&=&0
\\
\nonumber
\la \xi_\theta(\bbr, t) \xi_{\theta'}(\bbr',t') \ra &=& 2 \sigma \pp_\theta^2 \left[ \psi_\theta
\delta(\theta-\theta') 
\delta(t-t') \delta^2(\bbr-\bbr') \right]
\ .
\eeqn
where $\rho(\br) =(2\pi)^{-1}\int \dd \theta \psi_\theta(\br)$. 

We now follow exactly the steps in \cite{Alee_pre10} and expand $\psi_\theta(\br)$ as follows:
\beq
\psi_\theta(\bbr)= \alpha_0(\bbr)  +2\sum_{n\geq 1} \Big[ \alpha_n(\bbr) \cos (n \theta) +\beta_n (\bbr) \sin(n\theta)\Big]
\eeq
with
\beq
\alpha_n =\frac{1}{2\pi} \int_0^{2\pi} \dd \theta \cos (n\theta) \psi_\theta \sep
\beta_n = \frac{1}{2\pi}\int_0^{2\pi} \dd \theta \sin (n\theta) \psi_\theta
\ .
\eeq
In particular, $\alpha_0 =  \rho$,
$\alpha_n =\alpha_{-n}$ and $\beta_n =-\beta_{-n}$. 

In terms of these new fields, we have \cite{Alee_pre10}
\begin{subequations}
	\label{eq:fourierfield}
	\beqn
	\pp_t \alpha_n +  \frac{\pi \gamma}{2} \Big\{\pp_x \left[(\rho_M-\rho)(\alpha_{n+1}+\alpha_{n-1})\right] - \pp_y \left[(\rho_M-\rho)(\beta_{n+1}-\beta_{n-1})\right]\Big\}&=& -\sigma n^2 \alpha_n
	\\
	\pp_t \beta_n +\frac{\pi \gamma}{2} \Big\{ \pp_x\left[(\rho_M-\rho)(\beta_{n+1}+\beta_{n-1})\right] + \pp_y \left[(\rho_M-\rho)(\alpha_{n+1}-\alpha_{n-1})\right] \Big\}&=& -\sigma n^2\beta_n 
	\ .
	\eeqn
\end{subequations}
We have ignored the noise terms in the above equations and we will study their effects in the next section. However, we can see here that since the R.H.S. of (\ref{eq:fourierfield}) are always negative except for $n=0$, we have verified the hydrodynamic argument in the MT that all modes are massive except for the density mode $\rho = \alpha_0$.

	Since the ``mass'' of the $n$-th mode scales with $n^2$, the higher order modes are severely damped. If we now 
	truncate the set of equations by setting all $\alpha_n$ to zero for $n>1$, we recover Eq.~(11), where we also set all $\beta$ modes to zero for simplicity. 
	
	We now show how we can recover the non-equilibrium terms: $\nabla^4 \phi^2$ and $\nabla \cdot (\nabla \phi \nabla^2\phi)$ by incorporating higher order modes into the analysis. We again set all $\beta$ modes to zero for simplicity but keep track of $\alpha_n$ modes for all $n<2$. We will also set for $\pp_t \alpha_n$ to zero except for $\alpha_0$.
	
	We start with the $\alpha_2$ mode, which is
	\beq
	\label{Aeq:a2}
	\alpha_2 = A_2 \pp_x [(\rho_0 -\phi) \alpha_1]
	\eeq
	where $\rho_0 = \rho_M - \bar{\rho}$ with $\bar{\rho}$ being the average density, and  we have defined the constants $A_n \equiv \pi \gamma / (2 n^2 \sigma)$.
	
	The $\alpha_1$ is then
	\beq
	\label{Aeq:a1}
	\alpha_1 = A_1 \pp_x \Big[(\rho_0 -\phi) (\bar{\rho}+\phi+\alpha_2 )\Big]
	\ .
	\eeq
	
	In the above expressions, we should view both $\alpha_1$ and $\alpha_2$ as infinite series expansions of $\phi$ and its spatial derivatives. With this in mind, we can now see how the non-equilibrium term $\nabla^4 \phi^2$ emerges in the EOM. Its presence amounts a term of the form $\pp_x^3 \phi^2$ in the expansion of $\alpha_1$. To see how it arises, we first use \eq (\ref{Aeq:a1}) to ascertain that $\alpha_1$ has the term $-A_1 \pp_x \phi^2$, we then substitute that into  \eq (\ref{Aeq:a2}) to see that $\alpha_2$ has the term
	$ -A_1A_2 \rho_0 \pp_x^2 \phi^2$. Substituting this back into \eq (\ref{Aeq:a1}) leads to the desired $-A_1^2A_2 \rho_0^2 \pp_x^3 \phi^2$ term. 
	
	Similarly for the non-equilibrium $\nabla \cdot (\nabla \phi \nabla^2\phi)$, which corresponds to a term of the form $(\pp_x \phi) (\pp_x^2 \phi)$ in $\alpha_1$ in our one-dimensional representation (because, again, all $\beta$ modes are set to zero). We first use \eq (\ref{Aeq:a1}) to see that $\alpha_1$ has the term $-A_1 \alpha_2 \pp_x \phi$ (i). We then use  \eq (\ref{Aeq:a2}) to see that $\alpha_2$ has the term $A_2 \rho_0 \pp_x \alpha_1$ (ii). Now, \eq (\ref{Aeq:a1}) shows that $\alpha_1$ also the term $A_1 (\rho_0-\bar{\rho}) \pp_x\phi$ (iii). 
	Substituting (iii) into (ii) and then combining the expression with (i) gives the desired term $-A_1^2A_2 \rho_0(\rho_0-\bar{\rho}) (\pp_x \phi) (\pp_x^2 \phi)$.
	
	We note that although the above discussion focuses on a system that is varying along the $x$ direction only for simplicity (since all $\beta$ modes are set to zero), the fact that the system is rotational invariance implies that the terms obtained have to be rotationally invariant and thus do correspond to the non-equilibrium terms in their vectorial forms.

%

\section{Fluctuations-induced renormalizations of $a_1$ and $a_2$}
We have argued in the main text that the hydrodynamic EOM describing MIPS around the critical point is of the form:
\beq
\label{eq:red_main}
\pp_t \phi = \nabla^2 (a_1 \phi +a_2\phi^2 +a_3 \phi^3) +b \nabla^4 \phi + f
\ ,
\eeq
with  $\la| f(\bk,\omega )|^2 \ra =2D k^2$. Generally, all the coefficients in the EOM will be renormalized due to fluctuations and the nonlinearities. Here, we will demonstrate using graphical method that close to the critical point where $a_1$ and $a_2$ are small,  $a_1$ is renormalized upwards (i.e., $a_1$ becomes larger due to the fluctuations) while $a_2$ is renormalized downwards.

\begin{figure}
	\begin{center}
		\includegraphics[scale=.5]{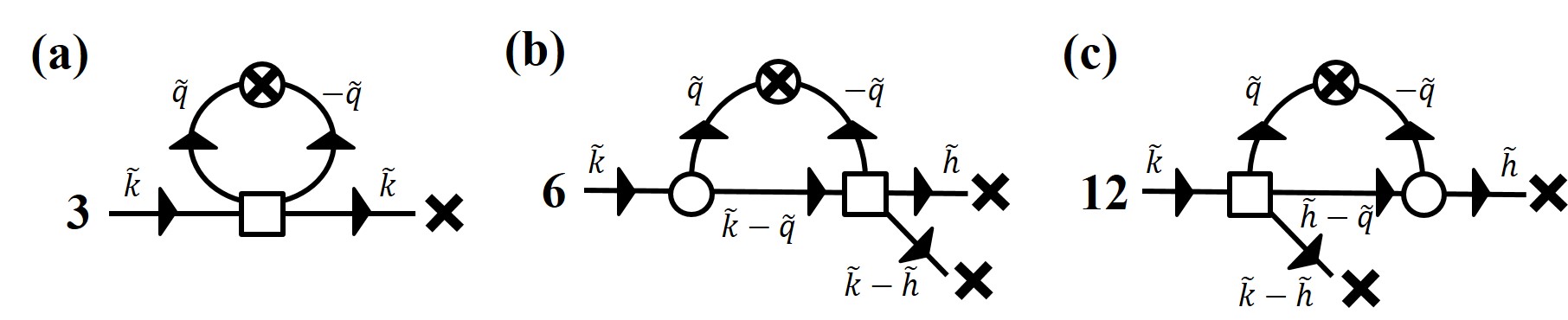}
	\end{center}
	\caption{Graphical contributions to $a_1$ (Fig.~a) and $a_2$ (Figs~b \& c) to linear order in $a_1$ and $a_2$ and to the one-loop level.
	}
	\label{fig:graphs}
\end{figure}

We start by spatially temporal and spatial Fourier transform the EOM to obtain:
\beq
\label{app:maineq}
\ii \omega \phi = -a_1 k^2 \phi - a_2k^2\int_{\tbq}\phi(\tbq)\phi(\tbk-\tbq) -a_3k^2 \int_{\tbq,\tbh}\phi(\tbq)\phi(\tbh)\phi(\tbk-\tbq-\tbh) -b k^4 \phi + f
\ . 
\eeq
The  bare propagator of the EOM is thus
\beq
G_0(\omega, \bk)= \frac{1}{\ii \omega + \Gamma(\bk)} 
\eeq
where 
\beq
\Gamma(\bk) =a_1k^2+b k^4
\ .
\eeq
Incorporating the fluctuations up to the one-loop level and to linear order in $a_1$ and $a_2$, the coefficient $a_1$ is modified by the graphical contribution shown in Fig.~\ref{fig:graphs}(a), while $a_2$ is modified by the sum of Figs~\ref{fig:graphs}(b) \& \ref{fig:graphs}(c). Note that since the model equation (\ref{app:maineq}) has both cubic and quartic terms, the graphs are identical to those discussed in \cite{Achen_njp15} (although the mathematical expressions of the propagator and vertices are of course different here), although we will only need a subset of them here.

\subsection{Graph (a)}

This graph represents an additional
contribution $\delta \left(\partial_t\phi\right)$ to $ \partial_t\phi$ given by:
\beqn
\delta\left( \partial_t\phi\right)&=&-3a_3 k^2 \phi(\tbk)\int_{\bq, \Omega} \frac{2D q^2}{\Gamma(\bq)^2 +\Omega^2}
\\ 
&=&-3 a_3 D k^2 \phi(\tbk)
\int_{\bq} \frac{q^2}{\Gamma(\bq) }
\\ 
&=&-3 a_3 D k^2 \phi(\tbk)
\int_{\bq} \frac{1}{bq^2+a_1}
\\
&=&-3 a_3 D  \frac{S_d}{(2\pi)^d} \frac{\Lambda^{d} \dd \ell}{b\Lambda^2 +a_1}k^2\phi(\tbk) 
\, .
\eeqn

Since this contribution is negative, this contribution increases $a_1$ in (\ref{app:maineq}).

\subsection{Graphs (b) \& (c)}

Graph (b) represents an additional
contribution $\delta \left(\partial_t\phi\right)$ to $ \partial_t\phi$ given by:
\beqn
\delta \left(\partial_t\phi\right)=6a_2a_3 k^2\int_{\tbh}\phi(\tbh)\phi(\tbk-\tbh)\int_{\bq, \Omega} \frac{ 2D q^2|\bk-\bq|^2}{(\Gamma(\bq)^2+\Omega^2)[\Gamma(\bk-\bq)+\ii(\omega-\Omega)]}\,,
\label{eq:Cgraphc}
\eeqn

Since there is already an explicit factor of $k^2$  in this expression, we can evaluate this graph by setting both $\bk$ and the external frequency $\omega$  to zero in the integrand of the integral over $\tbq$. Doing so, and in addition using
\beqn
\frac{1}{\Gamma(\bq)-\ii\Omega}=\frac{\Gamma(\bq)+\ii\Omega}{\Gamma(\bq)^2+\Omega^2}\,,
\label{propsimp}
\eeqn
gives, after dropping an integral of an odd function of $\Omega$ that vanishes,
\beqn
\delta\left( \partial_t\phi\right)=12a_2a_3 D k^2\int_{\tbh}\phi(\tbh)\phi(\tbk-\tbh)\int_{\bq, \Omega} \frac{q^4\Gamma(\bq)}{[\Gamma(\bq)^2+\Omega^2]^2}
\ .
\eeqn
The integral over frequency $\Omega$ and $\bq$ is readily evaluated, and is given by
\beqn
\int_{\bq, \Omega} \frac{q^4 \Gamma(\bq)}{[\Gamma(\bq)^2+\Omega^2]^2}=
\frac{1}{4} \int_\bq \frac{q^4}{\Gamma(\bq)^2}=
\frac{1}{4 b^2}\int_{\Lambda(1-\delta \ell)}^\Lambda  \frac{d^{d} q}{(2\pi)^{d-1}} \frac{1}{q^4}
=
\frac{1}{4b^2}{S_{d}\over\left(2\pi\right)^{d}}\Lambda^{d-4}d\ell\,.
\eeqn
We thus have from \eq (\ref{eq:Cgraphc})
\beqn
\label{app:graphb}
\delta \left(\partial_t\phi\right)&=&\frac{3 a_2a_3D}{b^2}{S_{d}\over\left(2\pi\right)^{d}}\Lambda^{d-4}d\ell \ k^2\int_{\tbh}\phi(\tbh)\phi(\tbk-\tbh)\, .
\eeqn

Graph (c) represents an additional
contribution $\delta \left(\partial_t\phi\right)$ to $ \partial_t\phi$ given by
\beqn
\delta \left(\partial_t\phi\right)=12 a_2a_3k^2 \int_{\tbh} \phi(\tbh)\phi(\tbk-\tbh)
\int_{\bq, \Omega} \frac{2Dq^2|\bh-\bq|^2}{[\Gamma(\bq)^2+\Omega^2][\Gamma(\bh-\bq)+\ii(\omega-\Omega)]}\ .
\label{eq:Agraphd}
\eeqn
Setting again $\bh$ and $\omega$ to zero, this is exactly of the form of (\ref{eq:Cgraphc}), we therefore have
\beqn
\label{app:graphc}
\delta \left(\partial_t\phi\right)&=&\frac{6 a_2a_3D}{b^2}{S_{d}\over\left(2\pi\right)^{d}}\Lambda^{d-4}d\ell \ k^2\int_{\tbh}\phi(\tbh)\phi(\tbk-\tbh)\, .
\eeqn

Since both contributions (\ref{app:graphb}) and (\ref{app:graphc}) are positive, they decrease $a_2$ in (\ref{app:maineq}).

\section{Simulation procedure}
\subsection{Model Outline}
Our model involves \(N\) polar ABPs whose dynamics evolve upon a planar hexagonal lattice (with lattice spacing $d=1$), subject to periodic boundary conditions and with steric interactions between particles only. We use an iterative update scheme for the evolution of particle position designed to mimic the dynamical Langevin equations for ABPs lacking alignment interactions. For a particle \(i\) at time \(t\) we implement the following update rules for its polarity angle \(\theta_i\): 
\begin{equation}
\theta_i(t) = \theta_{i}(t - 1) +  60^\circ \times \eta_i^{\theta}(t)
\ .
\end{equation}
where \(\eta_i^{\theta}\) is a rotational noise term generated by rounding a number drawn from a Gaussian distribution with zero mean and standard deviation \( \sigma \) to the nearest integer. 

And for the position vector \(\mathbf{r}_i\), we have
\begin{equation}
\mathbf{r}_i(t) = \mathbf{r}_i(t-1) + {\bf P^{\theta_i}_i}(t)
%
%
\label{eqn::position}
\end{equation}
where ${\bf P^{\theta_i}_i}(t)$ is chosen with probability 24/30 to be $\cos \theta_i(t) \hat{\bf x} +\sin \theta_i(t) \hat{\bf y}$ (representing the active motion), and with probability 6/30 to be a vector pointing to one of the six neighboring sites selected at random (representing the Brownian motion).

The iterative process proceeds by selecting a particle at random, and applying the previously described update scheme with an acceptance probability of unity if the destination lattice site is unoccupied, and probability zero if occupied. This obviously encoding the volume-exclusion effect implemented through a pairwise potential in molecular dynamics simulations. 

This scheme is similar to a kinetic Monte Carlo method proposed in \cite{Alevis2014}, only adapted to suit a lattice. Interestingly, the authors of this paper note that in models of ABPs which do not explicitly include an interaction term between particles, translational diffusion is necessary to instigate MIPS. This is corroborated by a recent study \cite{Awhitelam2018}, also conducted on-lattice, which claims that pre-constructed clusters will dissolve unless transverse motion is included in the simulation. It is noted that even within athermal ABP simulations, transverse motion is still possible due to repulsive force between colliding particles. Therefore, we have included translational diffusion within our simulations in order to mitigate any potential difficulty introduced by the omission of such a term.

When referring to the system size we note that due to the hexagonal geometry of the lattice the $y$-direction is scaled by a factor \(\sqrt{3}/2\). Meaning that the total size of the system in Euclidean space is actually \(\sqrt{3}L \times 6L\). However, throughout this text we work in units of rows and columns to avoid confusion; so a sub-box of size \(L \times L\) (with Euclidean area \(\sqrt{3}L^2/2\)) spans a total of \(L^2\) lattice sites.

\subsection{Finite Size Scaling at the Liquid-Gas Transition}
\begin{figure*}
	\hspace*{-0cm} 
	\includegraphics[scale=0.90]{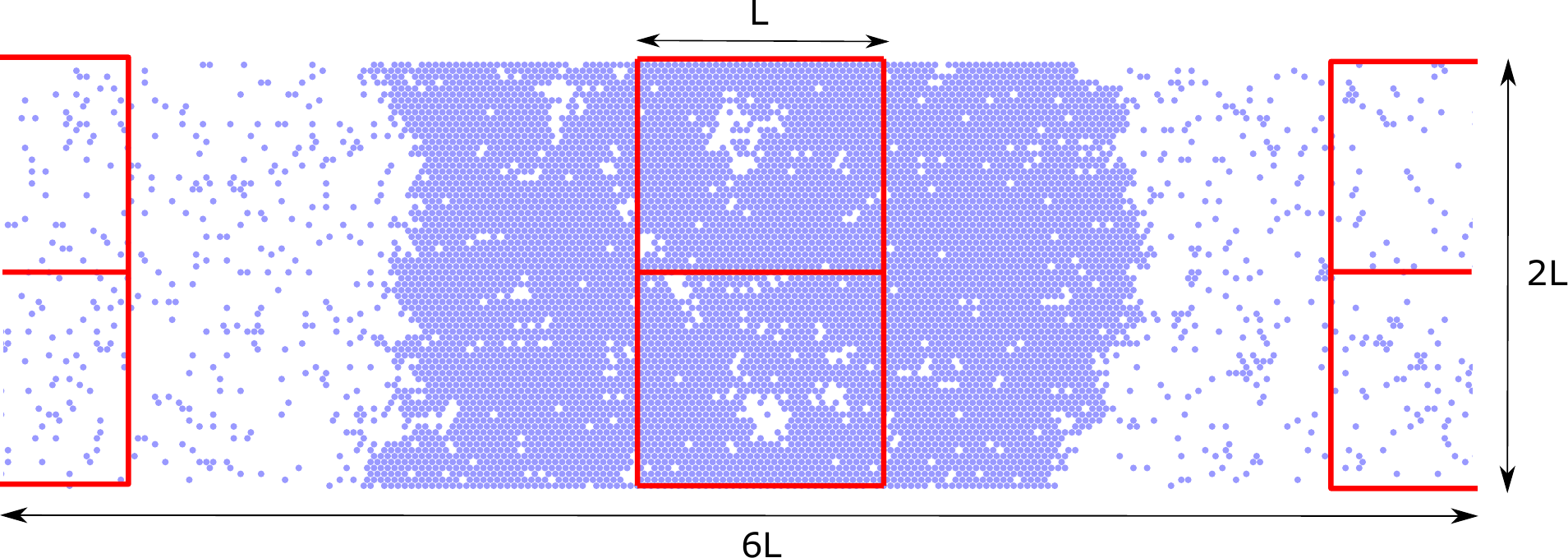}
	\caption{\textit{Sub-box sampling method.} We use an implementation of the sub-box sampling technique found in \cite{Asiebert2017}, adjusted to suit a hexagonal lattice. The total size of the system is fixed at \(2L \times 6L\). Note that the sub-box size is coupled to the total system size, in the original sub-box method, the sub-system size is varied for given total system size thus this technique eliminates the additional scaling parameter introduced by the sub-boxes \cite{Asiebert2017}.  A total of four sub-boxes of equal size \(L \times L\) are used, two for each corresponding phase. Sampling of the dense phase is fixed on the centre of mass of the system in the $x$-direction \(x_{COM}\). Likewise, sampling of the dilute phase is centred on \(x_{COM} + 3L\). This ensures that the sub-boxes are kept well away from the interface, therefore sampling the bulk densities of the co-existing phases accurately. By extracting time-series of density samples from a given simulation run we are able to construct steady-state distributions of the co-existing densities, which forms the basis for the calculation of the static exponents presented in the MT.}
	\label{sampling}
\end{figure*}
\subsubsection{Original Sub-Box Method for Equilibrium Liquid-Gas Systems}
A brief outline will now be given of the finite-size scaling (FSS) technique first developed in \cite{Abinder81} for the conventional Ising Model and subsequently extended to liquid-vapour systems. Rovere and colleagues \cite{Arovere90} adaptation is similar in its essence. Working in the canonical ensemble, a simulation box of linear dimension \( L \) is split into a grid of \((L/S)^d \) square sub-systems, where \( S \) is an integer and \( d \) the dimension of the system. The system has a fixed average density \(\langle \rho \rangle\), however, the sub-systems themselves possess a fluctuating particle number, thus it is useful to regard them as being in an effective quasi-grand canonical ensemble.  

We are interested in the moments of the average density distribution of the sub-systems:	
\begin{equation}
\langle \rho^k \rangle = \int P(\rho) \rho^k d\rho
\end{equation}
Where \(P(\rho)\) is the average probability distribution taken over all sub-boxes of the system. This is analogous to the distribution of the magnetisation per spin in the Ising model. From this, in principle we are able to extract all information regarding the nature of the transition. The zeroth moment is trivial due to normalisation of the distribution. In the case of the original sub-box method, the first moment is the average (fixed) density of the entire system. The second moment is the variance of the density fluctuations from this mean value.  
\begin{equation}
{\langle (\Delta\rho)^2 \rangle}_L = {\langle (\rho - \langle \rho \rangle)^2 \rangle}_L = L^{-d}\langle \rho \rangle^2 k_B T {K^{(L)}_T}
\end{equation}
Where the sub-script \(L\) indicates that we are average over simulation sub-boxes of a finite-size. Therefore, the isothermal compressibility of the distribution, related to the second moment through fluctuation analysis, is perturbed from its asymptotic value by some systematic error. \({K^{(L)}_T}\) attains it's true value \(K_T\) only in the limit \(L \rightarrow \infty\). 

In the homogeneous regime where \(L \gg \xi\), \(\xi\) being the characteristic length of density fluctuations. It is asserted that the distribution of sub-box density fluctuations will be a Gaussian centred on the average density of the system. 
\begin{equation}
p(\rho) \propto \exp\bigg(-\frac{(\rho - \langle \rho \rangle)^2}{2 \langle \rho \rangle ^2 k_B T K_T}\bigg)
\end{equation}
In the phase-separated regime, again where \(L \gg \xi\), the distribution instead takes on a superposition of two Gaussian peaks, each centred on the co-existing densities of the liquid and gas phases respectively. 
\begin{eqnarray}
p(\rho) \propto \frac{\rho_{liq} - \langle \rho \rangle}{\rho_{liq} - \rho_{gas}} \frac{1}{\rho_{gas} (K^{gas}_T)^{1/2}}   \exp\bigg(-\frac{(\rho - \rho_{gas})^2}{2 \rho_{gas}^2 k_B T K_T^{gas}}\bigg)  + \frac{\langle \rho \rangle - \rho_{gas}}{\rho_{liq} - \rho_{gas}} \frac{1}{\rho_{liq} (K^{liq}_T)^{1/2}}  \exp\bigg(-\frac{(\rho - \rho_{liq})^2}{2 \rho_{liq}^2 k_B T K_T^{liq}}\bigg)
\ ,
\label{twopeak}
\end{eqnarray}
where relative weight of two phases is attained using the well-known `lever rule'. A crucial point of note is that even in the thermodynamic limit  (\ref{twopeak}) is not exact. This is due to the fact that a significant proportion of the grid of sub-systems will contain the interface between the gas and liquid phase, necessitating the inclusion of additional terms in the free energy expansion. 

A celebrated tool in the numerical analysis of phase transitions is the so-called fourth order Binder cumulant: 
\begin{equation}
U_4^{(L)} = 1 - \frac{\langle (\Delta \rho)^4\rangle_L}{\langle (\Delta \rho)^2 \rangle_L^2}
\label{binder}
\end{equation}		
As first expounded in \cite{Abinder81}, this quantity is invariant with respect to system size at the critical point. In the case of the conventional Ising model the situation is relatively simple, due to the presence of the \(\phi \rightarrow -\phi\) symmetry of the order parameter. The situation is more complex in the case of the liquid-gas transition due to absence of such a symmetry, and so entails tuning two parameters, both the temperature and the average density of the system, in order to reach criticality. 

In order to identify the correct critical exponents we must tune the relevant system parameters (i.e. \( \sigma \)) such that we pass through the phase-separated regime, into the critical region and eventually into the well-mixed phase. In the critical region we expect that \( L \simeq \xi \) and for this reason the distribution becomes distinctly non-gaussian. It may be further postulated that in this region, the density distribution will obey a universal scaling function of the form \cite{Arovere90}: 
\begin{equation}
p_L(\rho) = L^{\beta/\nu} \tilde{p}[(\rho - \rho_c)L^{\beta/\nu}, (\langle \rho \rangle - \rho_c)|\epsilon|^{-\beta},L|\epsilon|^\nu]
\label{scaling}
\end{equation}	
As is put forward in \cite{Arovere90} the finite-size scaling relation for the susceptibility applies:
\begin{equation}
K_T^{(L)} = L^{\gamma/\nu}(\langle \rho \rangle k_B T)^{-1} f_2[(\langle \rho \rangle - \rho_c)|\epsilon|^{-\beta},L|\epsilon|^\nu]
\label{susceptibility}
\end{equation}	
where \(f_2\) is a scaling function and \(K_T^{(L)}\) is the finite-size compressibility, obtained through a calculation of the second moment of the density distribution for a given sub-system size \(L\).

\subsubsection{Adapted FSS Technique for Motility Induced Phase Separation}
\begin{figure}
	\hspace*{-0.0cm} 
	\includegraphics[scale=0.60,trim={0 0 0 0},clip]{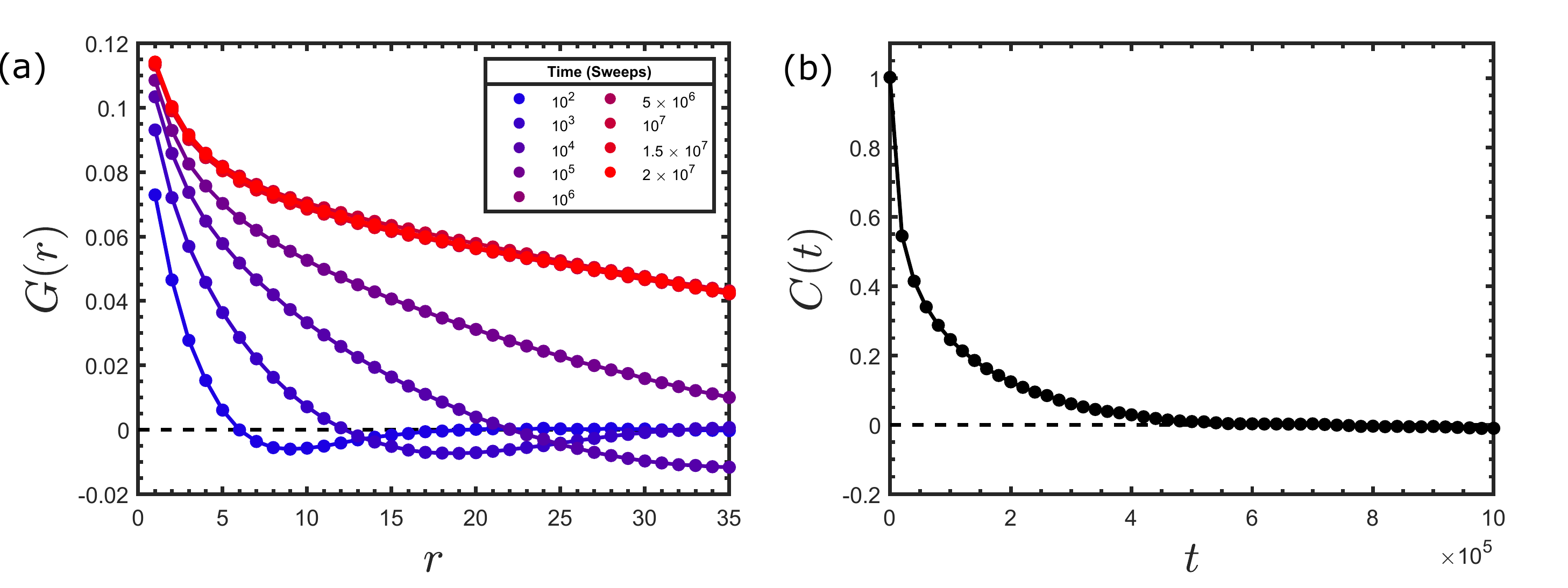}
	\caption{\textit{Determination of the equilibration and correlation time-scales around the critical point.} In each plot we work in a system size of \( L = 36\) and set the system parameters are set to their critical values. Plot (a) shows the results of the pair-correlation function \(G(r)\) averaged over one-hundred independent simulation runs at nine different time-lapses after initialisation. We take the static nature of \(G(r)\) with respect to time as an indication that the system has reached its steady-state. It is reasonable to assert that this appears to have occurred at around \(t \sim 10^6\) sweeps. Therefore, beginning sampling at time of \(t = 1 \times 10^7\) sweeps it is well-justified as the system is highly likely to have reach its steady state. Plot (b) shows the auto-correlation of the density \(C(t)\) of the two sub-box's fixed on the centre of mass of the system. By inspection we extract the time interval required to assert that two density samples taken from the same simulation run are approximately statistically independent as being equal to \(t_{relax} = 5 \times 10^5 \) sweeps.}
	\label{equilibration}	 
\end{figure}
In this section we describe our adaptation of the method put forward in \cite{Asiebert2017}, as an improvement to the scheme outlined above. In spirit it is identical to the original sub-box sampling technique, however, it mitigates a flaw which occurs due to the sampling of sub-systems containing interfaces. As shown in figure \ref{sampling} we are interested in the joint density distribution of the four sub-boxes, two of which are vertically adjacent and fixed upon the centre of mass of the system in the $x$-direction. And two further sub-boxes are then placed a distance \(3L\) in the $x$-direction, again also both vertically adjacent. For reasons given in \cite{Asiebert2017} this arrangement avoids the interfaces, and so focuses sampling upon the thermodynamic bulk in the phase co-existence regime.

One of the main problems in attempting to simulate the liquid-gas transition at its critical point is that the conserved nature of the order parameter (i.e. density fluctuations) introduces an finite-size effect into the system  that obfuscates access to the critical point. Here we utilise an elongated simulation box of aspect ratio 3:1 in order to force the steady-state phase separation into a slab like geometry. This has two primary advantages; firstly, it restricts density fluctuations at the interface to the $x$-direction only allowing the liquid phase to be sampled accurately by simply fixing the sampling boxes on the centre of mass in the x-direction. Secondly, and more importantly, as pointed out in \cite{Asiebert2017} it places an upper bound on the correlation length \(\xi\) to half (since we work under periodic boundary conditions) the shortest linear dimension of the system. This allows critical density fluctuations to be sampled in an effective grand canonical ensemble. This is not possible in an square simulation box since at the critical point the correlation length will inevitably span the entirety of the system, and therefore no sampling box with a fluctuating particle number - which also spans an area of size \(\xi^2\) - may be constructed.      

In our case, unlike the lattice gas, there is no particle hole symmetry. Therefore, the critical density is not known \textit{a priori} and must in principle also be systematically located along with the critical temperature. However, in this paper we assume that the deviation from the lattice gas case of \(\rho_c = 1/2\) is not so significant that it would preclude an estimate of the critical noise. Our procedure is as follows; we increment the strength of the rotational noise, beginning well within the phase separated regime, passing through the critical point and into the homogeneous region of the phase diagram. For each value of noise strength, three-hundred independent simulation runs were conducted of \(2 \times 10^7\) attempted moves per particle in length. 

\begin{figure}
	\hspace*{-0cm} 
	\includegraphics[scale=0.62,trim={0 0 0 0},clip]{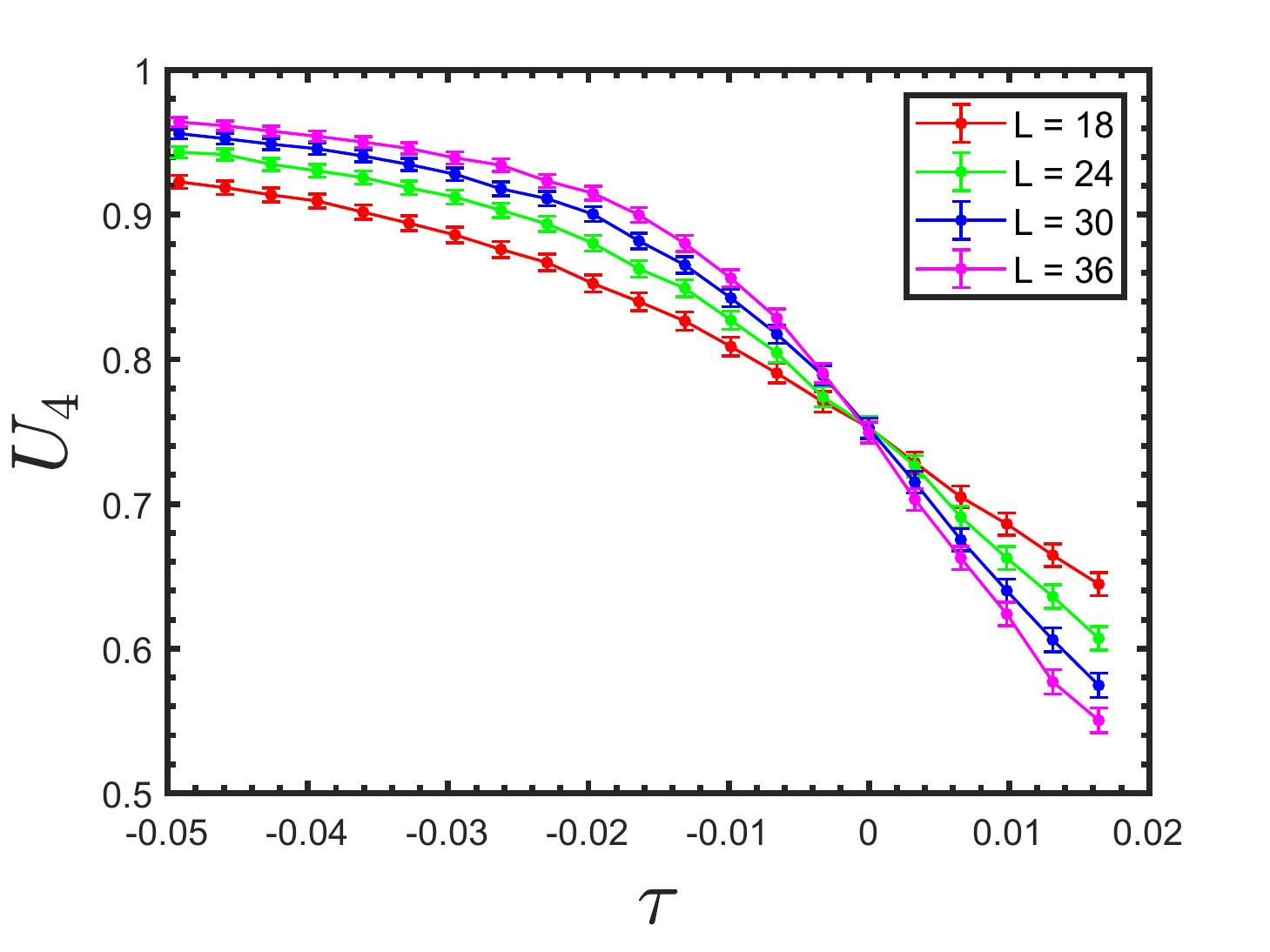}
	\caption{\textit{Binder cumulant \( U_4 \) in the critical region of the phase diagram.} Here we plot the Binder cumulant \(U_4\) as a function of the dimensionless distance to the critical noise for four different system sizes, the value of \( \sigma \) at which the curves intersect reveals the location of the asymptotic ($L \rightarrow \infty$) critical point, which is clearly identified as \( \sigma = 0.305  \).}
	\label{Binder_Cumulant}
\end{figure}

It is imperative that when computing the relevant statistical quantities (i.e. Binder cumulant) that the samples used are statistically independent. In order to ensure this, we computed the sub-box density auto-correlation function shown in \fig \ref{equilibration}, averaged over three-hundred simulation runs, and found that when within close proximity of the critical point (\( \tau = 0 \) in \fig \ref{Binder_Cumulant}), an average of \(5 \times 10^5\) particle sweeps was required to reduce the correlation to approximately zero.

Furthermore, we must also ensure that the system has reached its steady-state before sampling can begin. To estimate the time taken for this to occur, we deem the point at which the pair correlation function becomes static as indicative of the `equilibration' time. As can be observed from \fig \ref{equilibration}a) it is reasonable to assert that this time-scale is of the order \( \sim 10^6\) lattice sweeps. Thus in our simulations, sampling was commenced at \(1 \times 10^7\) lattice sweeps, ensuring with high probability that a steady state had been reached.  

\fig \ref{Binder_Cumulant} shows the Binder cumulant \(U_4\) calculated for four different sub-box sizes; with \(L \in \{18,24,30,36\}\). As in \cite{Asiebert2017} we calculate via
\begin{equation}
U_4 = \frac{\langle (\Delta \rho)^2\rangle^2_L}{\langle (\Delta \rho)^4 \rangle_L}
\end{equation}
where \( \Delta \rho = \rho - \langle \rho \rangle_L  \). The subscript \( \langle . \rangle_L\) representing an average over the four sub-boxes of finite size \(L\). Remarkably, despite only working in an average density \(\langle \rho \rangle\ = 1/2\), we see that the four curves coalesce at a definitive value of the critical rotational noise strength, namely \(\sigma_c = 0.305\).

In the asymptotic limit of $L \rightarrow \infty$, the correlation length \( \xi \) diverges at a precisely defined critical point. For a finite-size system, one can no longer speak of a critical point, but instead of a critical region of parameter space in which the correlation length becomes equal to its upper bound - the system size \(L\). Obviously, with increasing \(L\) this region will become increasingly localised, converging in the limit \(L \to \infty \). Replacing the correlation length \( \xi \) with the system size \( L \) forms the basis of finite-size scaling theory. As long as the relevant system parameters are tuned to values such that this condition is fulfilled; a calculation of the critical exponents becomes tractable. 

As demonstrated in the field-theoretic model, the critical density \( \rho_c \)  must be bounded below by a value of 1/2. Based on previous computational work on MIPS \cite{Asiebert2017} and similar studies of equilibrium liquid-gas phase separation \cite{Arovere93} it is reasonable to assume that \( \rho_c \) is slighter greater than this value, but much less that the upper bound of 3/4. Therefore, to estimate the critical noise, we work in an average density equal to the lower bound of 1/2 (equivalent to \( \rho_c \) in the lattice gas case). It is our assertion that this will not preclude an accurate estimate of the critical noise, this is justified by: (1) the convergence of the Binder cumulant \(U_4\) curves shown in figure \ref{Binder_Cumulant} for four different system sizes to a single point. If the deviation of the average density from its critical value was significant, then corrections due to the density term present in (\ref{scaling}) would prevent the curves in \fig \ref{Binder_Cumulant} from converging to a single point. (2) As mentioned in \cite{Arovere93} the estimate of the critical temperature (for liquid-vapour phase separation) seems insensitive to inaccuracy in the knowledge of \(\rho_c\).

In order to construct the phase diagram presented in \fig 2 of the MT, we note that due to the slab geometry of the phase separated steady state, the two sub-boxes in \fig \ref{sampling} centred on the system's COM will inevitably sample the bulk density of the liquid phase (likewise for vapour phase with the two sub-box's displaced \( 3L \) from the COM). Thus, by extracting a time series of density samples for three-hundred independent simulation runs for each value of \( \sigma \) shown in the phase diagram and computing the average, we are able to estimate the liquid-vapour binodal lines. The phase diagram in the MT uses data from the largest system size \(L = 36\). 

The critical density was estimated as in \cite{Asiebert2017} by averaging all 24000 independent density samples taken from three-hundred independent simulation runs from each of four sub-systems for the largest sub-system size \( L = 36 \) with the noise strength set to its critical value. This resulted in a rough estimate of \(\rho_c = 0.5223\), in line with the bounds dictated by our field-theoretic model.  For all calculations of exponents shown in \fig 3 of the main text the noise strength and average density were set to their estimated critical values (\(\rho_c = 0.5223\) and \(\sigma_c = 0.305\)).

\subsection{Calculation of Exponents}
\subsubsection{Static Exponents: \(\gamma\) \& \(\nu\)} 
Based on our hydrodynamic argument, only two independent exponents are required to identify the static universality class of the system under consideration. Typical finite-size scaling techniques of the kind first introduced in \cite{Abinder81} posit several possible routes to calculating various exponent ratios i.e. \(\alpha/\nu\), \(\beta/\nu\), \(\gamma/\nu\) etc. However, obtaining accurate estimates of the exponent \(\nu\) have posed a significant challenge. Here we use the Binder cumulant gradient, which should scale according to 
\begin{equation}
\bigg|\frac{\partial U_4}{\partial \tau}\bigg|_{\tau \simeq 0} \propto L^{1/\nu}
\end{equation}
when the finite-size system is tuned to the asymptotic critical parameters. 

As the critical point is approached, the values of \( U_4(\tau) \) begin to converge and their corresponding uncertainties increase. Referring to \fig \ref{Binder_Cumulant}, it is striking that our analysis provides a very good estimate of the critical noise at which all four points representing the values of \( U_4 \) become statistically indistinguishable. However, for the region surrounding this critical noise, at least two of the points (but not all) may be distinguished within their corresponding uncertainties. This introduces a minor difficulty in calculating the required gradients as we require that all points used in the intervals used in the calculation of gradients of each of the curves for different system size parameters \(L\) in \fig \ref{Binder_Cumulant} be statistically distinguishable. 

\begin{figure}
	\hspace*{-0cm} 
	\includegraphics[scale=0.58,trim={0 0 0 0},clip]{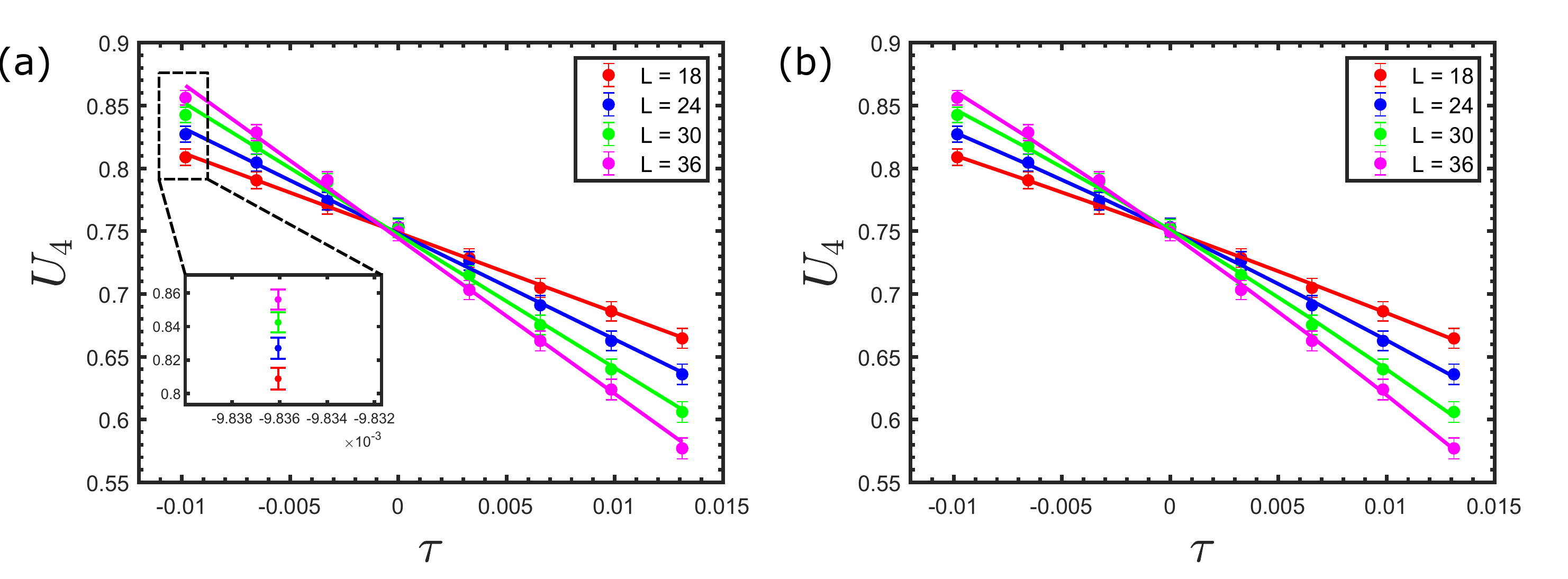}
	\caption{\textit{Estimation of Binder cumulant derivative at criticality}. We estimate the gradient of the Binder cumulant \( U_4 \) by fitting (a) a linear curve, and (b) a quadratic through the critical region. The `critical region' is defined as the interval enclosed by the first two values of \( \tau \) either side of the critical point for which each of the four system size dependent values of \(U_4\) become statistically distinguishable. The order of the polynomial used in the fitting does not effect the scaling result shown in the MT.}
	\label{fitting}
\end{figure}

Therefore, we estimate the gradient by fitting a polynomial through the region enclosed by the first two values of \( \tau \) either side of the critical noise for which each of the four system size dependent values of \( U_4 \) are distinguishable within their computed uncertainties. We show this in \fig \ref{fitting}. Importantly, the order of the polynomial used does not affect the scaling result displayed in the main text.

The exponent ratio $ \gamma/\nu $ was calculated using the finite-size scaling relation \( \chi_L \sim L^{\gamma/\nu} \). It is a well-known result from the analysis of equilibrium fluctuations \cite{Aroman97} that the isothermal compressibility \(\chi_L\) is given by
\begin{equation}
\chi_L = \frac{\langle N^2 \rangle_L - \langle N \rangle_L^2}{\langle N \rangle_L}
\end{equation} 
This is easily calculated from our time-series of sub-box density samples, again taking care to ensure that subsequent density samples are separated by a time interval of \( t_{relax} = 5 \times 10^5\) lattice sweeps. We estimate the uncertainty in the result using the standard propagation of error formulae, which, due to the large numbers of independent samples extracted also turns out to be negligible.  

\subsubsection{Dynamic Exponent $z$}
\begin{figure}
	\hspace*{-0cm} 
	\includegraphics[scale=0.62,trim={0 0 0 0},clip]{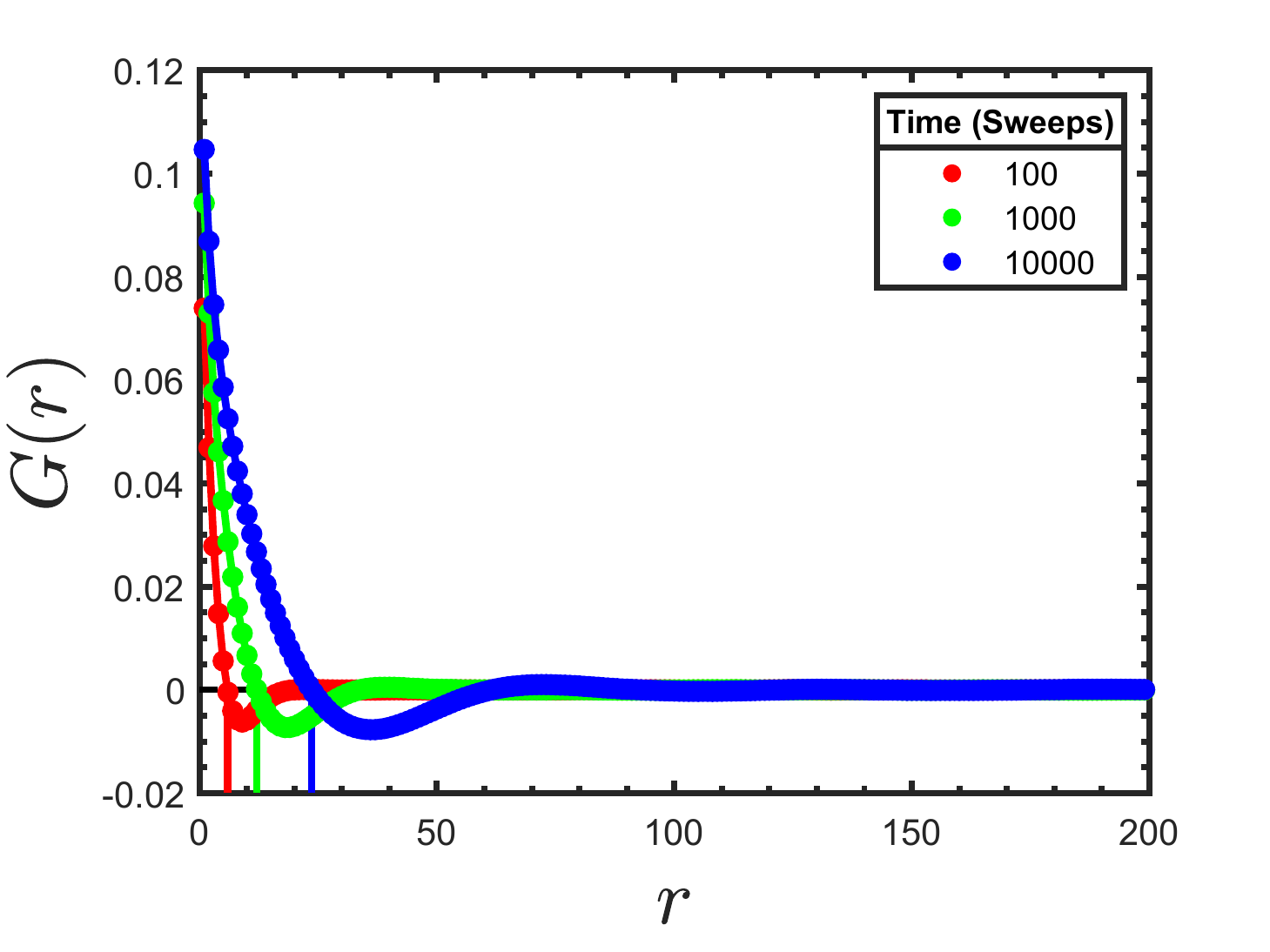}
	\caption{\textit{Determination of dynamical exponent \( z \).} Here we show the pair correlation function \( G(r) \) calculated at various time points after initialisation from a random state for system size \(L = 200 \). We may estimate the average length scale \( \ell (t) \) over which particle aggregates are correlated by ascertaining the first point of intersection of the curve with the $x$-axis, indicated above by the vertical lines.}
	\label{coarsening}
\end{figure}

\textit{Criticality} - In order to calculate the dynamical exponent, defined via the relation \(\tau_{char} \propto \xi^z\) we use an adapted version of the method presented in \cite{Aalexander94}. Starting from a completely disordered initial state, we calculated the coarsening length-scale  $\ell(t)$ of the system for the first four hundred-thousand sweeps of the simulation run. Exploiting the fact that this quantity should scale as $\ell(t) \propto t^{1/z}$. For conserved order parameter systems at criticality, phase separation will proceed through spinodal decomposition, providing a means by which to calculate a characteristic coarsening length-scale. It is reasonable to assert that an estimate of $\ell(t)$ is found from the value of distance in which the correlation function first becomes negative as this represents the average length-scale over which particles in the system are correlated over. An example of this process is shown in \fig \ref{coarsening}. 

For this computation we set the average density and noise strength to their critical values (\( \langle \rho_c \rangle = 0.5223\), \(\sigma_c = 0.305\)) as determined by the means described in the preceding section. The system size parameter was set to \(L = 200\). 
To obtain the curves shown in \fig \ref{coarsening} we averaged over 300 independent simulation runs using a simple linear interpolation between the last positive and first negative point of the correlation function to extract the corresponding value of the coarsening length-scale. We display three such correlation functions in figure \ref{coarsening}, illustrating the manner in which $\ell(t)$ increases with time. 

	\textit{Phase Separation} - As stated in the main text, due to the emergence of the Gibbs-Thompson relation at the interface we expect that for phase separation \(\ell(t)\) will scale as \( t^{1/3} \) in accordance with the Lifshitz-Slyozov scaling law. Therefore, we also repeated the numerical analysis described above deep within the phase separated regime. In this case we use a hexagonal lattice in a square simulation box of size \(1000 \times 1000\) lattice sites, with average density \(\langle \rho \rangle = 0.25\) and  noise strength \(\sigma = 0.24\).  Again, we calculate \(\ell(t)\) over the first four hundred-thousand particle sweeps and average over 100 independent simulation runs initialised from a random state.
	
	As is noted in \cite{Agonella2015} the system will go through an initial transient period of clustering not in accordance with the power law scaling for coarsening; this appears to occur when \(\ell(t)\) is less than the persistence length of a single active Brownian particle, given by \(l_p = v_0 \tau = v_0/D_r\). The relatively large simulation run-time and number of particles (\(n = 250000\)) used allow us to bypass this effect and we recover good correspondence with the expected analytical results in the long-time limit. Notably, a fitting of the coarsening simulation results gives an exponent of \(\sim 0.34\) which is closer to the 1/3 scaling than reported in other studies using two-dimensional ABPs namely 0.272 in \cite{Aredner2013} and 0.28 in \cite{Astenhammar2014}. 

\subsubsection{Error estimation in \fig 3 of the MT}

{\it \fig 3a.} For each simulation run we allow the system to reach its steady-state and sample at intervals of the correlation time. Over three-hundred independent simulation runs we collect a total of \(N = 24000\) independent density samples (8000 for each of the four sub-systems). We estimate the error in the mean of this distribution through the well-known formula
\begin{equation}
\mathrm{error}  = \sqrt{\frac{\sum_1^N (\delta \rho_i)^2}{(N-1)N}}
\end{equation}
where \(\delta \rho_i  = \rho_i - (1/N) \sum_i^N \rho_i \). And likewise for the error in other higher order moments \(\langle \delta \rho^2 \rangle\) etc. Using the standard formulae for the propagation of uncertainties we calculate the error in the susceptibility \(\chi_L\) at the critical noise; which turns out to be negligible in comparison to the size of the markers in \fig 3 in the MT.

{\it \fig 3b.} The uncertainty in the gradient of the binder cumulant \( U_4 \) at criticality was estimated by extracting the 95 percent confidence intervals in the coefficients obtained through a linear least squares fitting of \( U_4(\tau) \) in the critical region with a polynomial function. For a linear fit \(y = ax + b\) this is obviously the uncertainty in \(a\). While for a quadratic fit  \( y = cx^2 + dx + e \) (with the gradient evaluated a  \( \tau = 0 \)) this would be related to the uncertainty in \(d\). Again we found that the size of the error was comparable to the size of the markers in \fig 3 of the MT.

{\it \fig 3c.} The connected pair correlation function \( G(r) \) is calculated for times between 1000 and 400000 particle sweeps for three-hundred independent simulation runs. For each time \(t\) we can calculate an average \(\tilde{G}(r)\) and estimate the corresponding uncertainty in the usual manner. To estimate the uncertainty in the coarsening length \( \ell(t) \) we simply use the difference between the maximum and minimum possible results allowed for by the uncertainties in the two points either side of the first intercept of \(\tilde{G}(r)\) with the \(r\) axis. Again due to the large system size and large number of independent runs the errors were negligible.


\begin{thebibliography}{30}
	\expandafter\ifx\csname natexlab\endcsname\relax\def\natexlab#1{#1}\fi
	\expandafter\ifx\csname bibnamefont\endcsname\relax
	\def\bibnamefont#1{#1}\fi
	\expandafter\ifx\csname bibfnamefont\endcsname\relax
	\def\bibfnamefont#1{#1}\fi
	\expandafter\ifx\csname citenamefont\endcsname\relax
	\def\citenamefont#1{#1}\fi
	\expandafter\ifx\csname url\endcsname\relax
	\def\url#1{\texttt{#1}}\fi
	\expandafter\ifx\csname urlprefix\endcsname\relax\def\urlprefix{URL }\fi
	\providecommand{\bibinfo}[2]{#2}
	\providecommand{\eprint}[2][]{\url{#2}}
	
	\bibitem[{\citenamefont{Marchetti et~al.}(2013)\citenamefont{Marchetti, Joanny,
			Ramaswamy, Liverpool, Prost, Rao, and Simha}}]{marchetti_rmp13}
	\bibinfo{author}{\bibfnamefont{M.~C.} \bibnamefont{Marchetti}},
	\bibinfo{author}{\bibfnamefont{J.~F.} \bibnamefont{Joanny}},
	\bibinfo{author}{\bibfnamefont{S.}~\bibnamefont{Ramaswamy}},
	\bibinfo{author}{\bibfnamefont{T.~B.} \bibnamefont{Liverpool}},
	\bibinfo{author}{\bibfnamefont{J.}~\bibnamefont{Prost}},
	\bibinfo{author}{\bibfnamefont{M.}~\bibnamefont{Rao}}, \bibnamefont{and}
	\bibinfo{author}{\bibfnamefont{R.~A.} \bibnamefont{Simha}}, 
	{\it Hydrodynamics of soft active matter},
	\bibinfo{journal}{Reviews of Modern Physics} \textbf{\bibinfo{volume}{85}},
	\bibinfo{pages}{1143} (\bibinfo{year}{2013}).
	
	\bibitem[{\citenamefont{Vicsek et~al.}(1995)\citenamefont{Vicsek, Czir{\'{o}}k,
			Ben-Jacob, Cohen, and Shochet}}]{Vicsek1995}
	\bibinfo{author}{\bibfnamefont{T.}~\bibnamefont{Vicsek}},
	\bibinfo{author}{\bibfnamefont{A.}~\bibnamefont{Czir{\'{o}}k}},
	\bibinfo{author}{\bibfnamefont{E.}~\bibnamefont{Ben-Jacob}},
	\bibinfo{author}{\bibfnamefont{I.}~\bibnamefont{Cohen}}, \bibnamefont{and}
	\bibinfo{author}{\bibfnamefont{O.}~\bibnamefont{Shochet}},
	{\it Novel Type of Phase Transition in a System of Self-Driven Particles},
	\bibinfo{journal}{Physical Review Letters} \textbf{\bibinfo{volume}{75}},
	\bibinfo{pages}{1226} (\bibinfo{year}{1995}).
	
	\bibitem[{\citenamefont{Toner and Tu}(1995)}]{toner_prl95}
	\bibinfo{author}{\bibfnamefont{J.}~\bibnamefont{Toner}} \bibnamefont{and}
	\bibinfo{author}{\bibfnamefont{Y.}~\bibnamefont{Tu}},
	{\it Long-Range Order in a Two-Dimensional Dynamical XY Model: How Birds Fly Together},
	\bibinfo{journal}{Physical Review Letters} \textbf{\bibinfo{volume}{75}},
	\bibinfo{pages}{4326} (\bibinfo{year}{1995}).
	
	\bibitem[{\citenamefont{Toner and Tu}(1998)}]{toner_pre98}
	\bibinfo{author}{\bibfnamefont{J.}~\bibnamefont{Toner}} \bibnamefont{and}
	\bibinfo{author}{\bibfnamefont{Y.}~\bibnamefont{Tu}},
	{\it Flocks, herds, and schools: A quantitative theory of flocking},
	\bibinfo{journal}{Physical Review E} \textbf{\bibinfo{volume}{58}},	
	\bibinfo{pages}{4828} (\bibinfo{year}{1998}).
	
	\bibitem[{\citenamefont{Tailleur and Cates}(2008)}]{tailleur_prl08}
	\bibinfo{author}{\bibfnamefont{J.}~\bibnamefont{Tailleur}} \bibnamefont{and}
	\bibinfo{author}{\bibfnamefont{M.~E.} \bibnamefont{Cates}},
	{\it Statistical mechanics of interacting run-and-tumble bacteria},
	\bibinfo{journal}{Physical review letters} \textbf{\bibinfo{volume}{100}},
	\bibinfo{pages}{218103} (\bibinfo{year}{2008}).
	
	\bibitem[{\citenamefont{Fily and Marchetti}(2012)}]{fily_prl12}
	\bibinfo{author}{\bibfnamefont{Y.}~\bibnamefont{Fily}} \bibnamefont{and}
	\bibinfo{author}{\bibfnamefont{M.~C.} \bibnamefont{Marchetti}},
	{\it Athermal Phase Separation of Self-Propelled Particles with No Alignment},
	\bibinfo{journal}{Physical Review Letters} \textbf{\bibinfo{volume}{108}},
	\bibinfo{pages}{235702} (\bibinfo{year}{2012}).
	
	\bibitem[{\citenamefont{Redner et~al.}(2013)\citenamefont{Redner, Hagan, and
			Baskaran}}]{redner_prl13}
	\bibinfo{author}{\bibfnamefont{G.~S.} \bibnamefont{Redner}},
	\bibinfo{author}{\bibfnamefont{M.~F.} \bibnamefont{Hagan}}, \bibnamefont{and}
	\bibinfo{author}{\bibfnamefont{A.}~\bibnamefont{Baskaran}},
	{\it 	Structure and Dynamics of a Phase-Separating Active Colloidal Fluid},
	\bibinfo{journal}{Physical Review Letters} \textbf{\bibinfo{volume}{110}},
	\bibinfo{pages}{055701} (\bibinfo{year}{2013}).
	
	\bibitem[{\citenamefont{Cates and Tailleur}(2015)}]{Cates2015}
	\bibinfo{author}{\bibfnamefont{M.~E.} \bibnamefont{Cates}} \bibnamefont{and}
	\bibinfo{author}{\bibfnamefont{J.}~\bibnamefont{Tailleur}},
	{\it Motility-Induced Phase Separation},
	\bibinfo{journal}{Annual Review of Condensed Matter Physics}
	\textbf{\bibinfo{volume}{6}}, \bibinfo{pages}{219} (\bibinfo{year}{2015}).
	
	\bibitem[{\citenamefont{Cates and Tjhung}(2018)}]{cates_annrev18}
	\bibinfo{author}{\bibfnamefont{M.~E.} \bibnamefont{Cates}} \bibnamefont{and}
	\bibinfo{author}{\bibfnamefont{E.}~\bibnamefont{Tjhung}},
	{\it Theories of binary fluid mixtures: from phase-separation kinetics to active emulsions},
	\bibinfo{journal}{Journal of Fluid Mechanics} \textbf{\bibinfo{volume}{836}},
	\bibinfo{pages}{P1} (\bibinfo{year}{2018}).
	
	\bibitem[{\citenamefont{Chen et~al.}(2016)\citenamefont{Chen, Lee, and
			Toner}}]{chen_natcomm16}
	\bibinfo{author}{\bibfnamefont{L.}~\bibnamefont{Chen}},
	\bibinfo{author}{\bibfnamefont{C.~F.} \bibnamefont{Lee}}, \bibnamefont{and}
	\bibinfo{author}{\bibfnamefont{J.}~\bibnamefont{Toner}},
	{\it Mapping two-dimensional polar active fluids to two-dimensional soap and one-dimensional sandblasting},
	\bibinfo{journal}{Nature Communications} \textbf{\bibinfo{volume}{7}},
	\bibinfo{pages}{12215} (\bibinfo{year}{2016}).
	
	\bibitem[{\citenamefont{Chen et~al.}(2018)\citenamefont{Chen, Lee, and
			Toner}}]{chen_a18}
	\bibinfo{author}{\bibfnamefont{L.}~\bibnamefont{Chen}},
	\bibinfo{author}{\bibfnamefont{C.~F.} \bibnamefont{Lee}}, \bibnamefont{and}
	\bibinfo{author}{\bibfnamefont{J.}~\bibnamefont{Toner}},
	{\it Squeezed in three dimensions, moving in two: Hydrodynamic theory of 3D incompressible easy-plane polar active fluids}, Physical Review E  {\bf 98}, 040602 (2018)
	
	\bibitem[{\citenamefont{Gr{\'{e}}goire and Chat{\'{e}}}(2004)}]{gregoire_prl04}
	\bibinfo{author}{\bibfnamefont{G.}~\bibnamefont{Gr{\'{e}}goire}}
	\bibnamefont{and}
	\bibinfo{author}{\bibfnamefont{H.}~\bibnamefont{Chat{\'{e}}}},
	{\it Onset of Collective and Cohesive Motion},
	\bibinfo{journal}{Physical Review Letters} \textbf{\bibinfo{volume}{92}},
	\bibinfo{pages}{025702} (\bibinfo{year}{2004}).
	
	
	
	\bibitem[{\citenamefont{Toner}(2012)}]{toner_pre12}
	\bibinfo{author}{\bibfnamefont{J.}~\bibnamefont{Toner}},
	{\it Reanalysis of the hydrodynamic theory of fluid, polar-ordered flocks},
	\bibinfo{journal}{Physical Review E} \textbf{\bibinfo{volume}{86}},
	\bibinfo{pages}{031918} (\bibinfo{year}{2012}).
	
	
	
	\bibitem[{\citenamefont{Wittkowski et~al.}(2014)\citenamefont{Wittkowski,
			Tiribocchi, Stenhammar, Allen, Marenduzzo, and Cates}}]{wittkowski_natcomm14}
	\bibinfo{author}{\bibfnamefont{R.}~\bibnamefont{Wittkowski}},
	\bibinfo{author}{\bibfnamefont{A.}~\bibnamefont{Tiribocchi}},
	\bibinfo{author}{\bibfnamefont{J.}~\bibnamefont{Stenhammar}},
	\bibinfo{author}{\bibfnamefont{R.~J.} \bibnamefont{Allen}},
	\bibinfo{author}{\bibfnamefont{D.}~\bibnamefont{Marenduzzo}},
	\bibnamefont{and} \bibinfo{author}{\bibfnamefont{M.~E.} \bibnamefont{Cates}},
	{\it Scalar $\phi^4$ field theory for active-particle phase separation},
	\bibinfo{journal}{Nature Communications} \textbf{\bibinfo{volume}{5}},
	\bibinfo{pages}{4351} (\bibinfo{year}{2014}).
	
	\bibitem[{\citenamefont{Tjhung et~al.}(2018)\citenamefont{Tjhung, Nardini, and
			Cates}}]{Tjhung2018}
	\bibinfo{author}{\bibfnamefont{E.}~\bibnamefont{Tjhung}},
	\bibinfo{author}{\bibfnamefont{C.}~\bibnamefont{Nardini}}, \bibnamefont{and}
	\bibinfo{author}{\bibfnamefont{M.~E.} \bibnamefont{Cates}},
	{\it Cluster Phases and Bubbly Phase Separation in Active Fluids: Reversal of the Ostwald Process},
	Physical Review X {\bf 8}, 031080 (2018).
	
	\bibitem[{\citenamefont{Cardy}(1996)}]{Cardy1996}
	\bibinfo{author}{\bibfnamefont{J.}~\bibnamefont{Cardy}},
	\emph{\bibinfo{title}{{Scaling and Renormalization in Statistical Physics}}}
	(\bibinfo{publisher}{Cambridge University Press}, \bibinfo{year}{1996}).
	
	\bibitem[{\citenamefont{Kardar}(2007)}]{Kardar2007}
	\bibinfo{author}{\bibfnamefont{M.}~\bibnamefont{Kardar}},
	\emph{\bibinfo{title}{{Statistical Physics of Fields}}}
	(\bibinfo{publisher}{Cambridge University Press}, \bibinfo{year}{2007}).
	
	\bibitem[{\citenamefont{Hohenberg and Halperin}(1977)}]{hohenberg_rmp77}
	\bibinfo{author}{\bibfnamefont{P.~C.} \bibnamefont{Hohenberg}}
	\bibnamefont{and} \bibinfo{author}{\bibfnamefont{B.~I.} \bibnamefont{Halperin},
		{\it Theory of dynamic critical phenomena},}, \bibinfo{journal}{Reviews of Modern Physics}
	\textbf{\bibinfo{volume}{49}}, \bibinfo{pages}{435} (\bibinfo{year}{1977}).
	
	\bibitem[{\citenamefont{Lee}(2013)}]{lee_njp13}
	\bibinfo{author}{\bibfnamefont{C.~F.} \bibnamefont{Lee}},
	{\it Active particles under confinement: Aggregation at the wall and gradient formation inside a channel}, \bibinfo{journal}{New
		Journal of Physics} \textbf{\bibinfo{volume}{15}}, \bibinfo{pages}{055007}
	(\bibinfo{year}{2013}).
	
	\bibitem[{\citenamefont{Lef{\`{e}}vre and Biroli}(2007)}]{lefevre_jstatmech07}
	\bibinfo{author}{\bibfnamefont{A.}~\bibnamefont{Lef{\`{e}}vre}}
	\bibnamefont{and} \bibinfo{author}{\bibfnamefont{G.}~\bibnamefont{Biroli}},
	{\it Dynamics of interacting particle systems: stochastic process and field theory},
	\bibinfo{journal}{Journal of Statistical Mechanics: Theory and Experiment}
	\textbf{\bibinfo{volume}{2007}}, \bibinfo{pages}{P07024}
	(\bibinfo{year}{2007}).
	
	\bibitem[{\citenamefont{Lee}(2010)}]{lee_pre_CM10}
	\bibinfo{author}{\bibfnamefont{C.~F.} \bibnamefont{Lee}},
	{\it Fluctuation-induced collective motion: A single-particle density analysis},
	\bibinfo{journal}{Physical Review E} \textbf{\bibinfo{volume}{81}}, 031125 (\bibinfo{year}{2010}).
	
	\bibitem[{\citenamefont{Bertin et~al.}(2006)\citenamefont{Bertin, Droz, and
			Gr{\'{e}}goire}}]{bertin_pre06}
	\bibinfo{author}{\bibfnamefont{E.}~\bibnamefont{Bertin}},
	\bibinfo{author}{\bibfnamefont{M.}~\bibnamefont{Droz}}, \bibnamefont{and}
	\bibinfo{author}{\bibfnamefont{G.}~\bibnamefont{Gr{\'{e}}goire}},
	{\it Boltzmann and hydrodynamic description for self-propelled particles},
	\bibinfo{journal}{Physical Review E} \textbf{\bibinfo{volume}{74}},
	\bibinfo{pages}{022101} (\bibinfo{year}{2006}).
	
	\bibitem[{\citenamefont{Peruani et~al.}(2008)\citenamefont{Peruani, Deutsch,
			and B{\"{a}}r}}]{Peruani2008}
	\bibinfo{author}{\bibfnamefont{F.}~\bibnamefont{Peruani}},
	\bibinfo{author}{\bibfnamefont{A.}~\bibnamefont{Deutsch}}, \bibnamefont{and}
	\bibinfo{author}{\bibfnamefont{M.}~\bibnamefont{B{\"{a}}r}},
	{\it A mean-field theory for self-propelled particles interacting by velocity alignment mechanisms},
	\bibinfo{journal}{The European Physical Journal - Special Topics}
	\textbf{\bibinfo{volume}{157}}, \bibinfo{pages}{111} (\bibinfo{year}{2008}).
	
	\bibitem[{\citenamefont{Bertin et~al.}(2009)\citenamefont{Bertin, Droz, and
			Gregoire}}]{bertin_jpa09}
	\bibinfo{author}{\bibfnamefont{E.}~\bibnamefont{Bertin}},
	\bibinfo{author}{\bibfnamefont{M.}~\bibnamefont{Droz}}, \bibnamefont{and}
	\bibinfo{author}{\bibfnamefont{G.}~\bibnamefont{Gr\'{e}goire}},
	{\it Hydrodynamic equations for self-propelled particles: microscopic derivation and stability analysis},
	\bibinfo{journal}{Journal of Physics A: Mathematical and Theoretical}
	\textbf{\bibinfo{volume}{42}}, \bibinfo{pages}{445001}
	(\bibinfo{year}{2009}).
	
	\bibitem{solon_pre18}
	A.P.~Solon, J.~Stenhammar, M.E.~Cates,Y.~Kafri, andJ.~Tailleur, \emph{Generalized thermodynamics of phase equilibria in scalar active matter}, Phys. Rev. E {\bf 97} 020602 (2018).
	
	
	
	
	\bibitem[{\citenamefont{Siebert et~al.}(2017)\citenamefont{Siebert, Dittrich,
			Schmid, Binder, Speck, and Virnau}}]{siebert_a17}
	\bibinfo{author}{\bibfnamefont{J.~T.} \bibnamefont{Siebert}},
	\bibinfo{author}{\bibfnamefont{F.}~\bibnamefont{Dittrich}},
	\bibinfo{author}{\bibfnamefont{F.}~\bibnamefont{Schmid}},
	\bibinfo{author}{\bibfnamefont{K.}~\bibnamefont{Binder}},
	\bibinfo{author}{\bibfnamefont{T.}~\bibnamefont{Speck}}, \bibnamefont{and}
	\bibinfo{author}{\bibfnamefont{P.}~\bibnamefont{Virnau}},
	{\it Critical behavior of active Brownian particles},
	\bibinfo{journal}{Physical Review E} \textbf{\bibinfo{volume}{98}},
	\bibinfo{pages}{030601} (\bibinfo{year}{2018}).
	
	\bibitem[{\citenamefont{Landau and Binder}(2014)}]{landau_b2014}
	\bibinfo{author}{\bibfnamefont{D.~P.} \bibnamefont{Landau}} \bibnamefont{and}
	\bibinfo{author}{\bibfnamefont{K.}~\bibnamefont{Binder}},
	\emph{\bibinfo{title}{{A Guide to Monte Carlo Simulations in Statistical
				Physics}}} (\bibinfo{publisher}{Cambridge University Press}, 2014).
	
	\bibitem[{\citenamefont{Alexander et~al.}(1994)\citenamefont{Alexander, Huse,
			and Janowsky}}]{alexander_prb94}
	\bibinfo{author}{\bibfnamefont{F.~J.} \bibnamefont{Alexander}},
	\bibinfo{author}{\bibfnamefont{D.~A.} \bibnamefont{Huse}}, \bibnamefont{and}
	\bibinfo{author}{\bibfnamefont{S.~A.} \bibnamefont{Janowsky}},
	{\it Dynamical scaling and decay of correlations for spinodal decomposition at $T_c$},
	\bibinfo{journal}{Physical Review B} \textbf{\bibinfo{volume}{50}},
	\bibinfo{pages}{663} (\bibinfo{year}{1994}).
	
	\bibitem[{\citenamefont{Lee}(2017)}]{lee_softmatter17}
	\bibinfo{author}{\bibfnamefont{C.~F.} \bibnamefont{Lee}},
	{\it Interface stability, interface fluctuations, and the Gibbs–Thomson relationship in motility-induced phase separations},
	\bibinfo{journal}{Soft Matter} \textbf{\bibinfo{volume}{13}},
	\bibinfo{pages}{376} (\bibinfo{year}{2017}).
	
	\bibitem{solon_njp18}
	A.P. Solon, J. Stenhammar, M.E. Cates, Y. Kafri, and J. Tailleur,  {\it Generalized thermodynamics of motility-induced phase separation: phase equilibria, Laplace pressure, and change of ensembles} New J. Phys. {\bf 20} 075001 (2018)
	
	
	
	
	
	\bibitem[{\citenamefont{Chen et~al.}(2015)\citenamefont{Chen, Toner, and
			Lee}}]{chen_njp15}
	\bibinfo{author}{\bibfnamefont{L.}~\bibnamefont{Chen}},
	\bibinfo{author}{\bibfnamefont{J.}~\bibnamefont{Toner}}, \bibnamefont{and}
	\bibinfo{author}{\bibfnamefont{C.~F.} \bibnamefont{Lee}},
	{\it Critical phenomenon of the order-disorder transition in incompressible active fluids},
	\bibinfo{journal}{New Journal of Physics} \textbf{\bibinfo{volume}{17}},
	\bibinfo{pages}{042002} (\bibinfo{year}{2015}).
	
	\bibitem[{\citenamefont{Caballero et~al.}(2018)\citenamefont{Caballero, Nardini, and
			Cates}}]{caballero_2018}
	\bibinfo{author}{\bibfnamefont{F.}~\bibnamefont{Caballero}},
	\bibinfo{author}{\bibfnamefont{C.}~\bibnamefont{Nardini}}, \bibnamefont{and}
	\bibinfo{author}{\bibfnamefont{M.}~\bibnamefont{Cates}},
	{\it From bulk to microphase separation in scalar active matter: a perturbative renormalization group analysis},
	\bibinfo{journal}{J. Stat. Mech.}, \bibinfo{pages}{123208} (\bibinfo{year}{2018}).
	
	\bibitem[{\citenamefont{Thompson et~al.}(2011)\citenamefont{Caballero, Nardini, and
			Cates}}]{thompson_2011}
	\bibinfo{author}{\bibfnamefont{A.~G} \bibnamefont{Thompson}},
	\bibinfo{author}{\bibfnamefont{J.}~\bibnamefont{Tailleur}}, 
	\bibinfo{author}{\bibfnamefont{M.~E} \bibnamefont{Cates}},
	\bibinfo{author}{\bibfnamefont{R.~A} \bibnamefont{Blythe}},
	{\it Lattice models of nonequilibrium bacterial dynamics},
	\bibinfo{journal}{Journal of Statistical Mechanics: Theory and Experiment},
	\bibinfo{pages}{P02029} (\bibinfo{year}{2011}).
	
	\bibitem[{\citenamefont{Soto et~al.}(2014)\citenamefont{Soto and Goldstein}}]{soto_2014}
	\bibinfo{author}{\bibfnamefont{R.}~ \bibnamefont{Soto}} \bibnamefont{and}
	\bibinfo{author}{\bibfnamefont{R.}~\bibnamefont{Goldstein}}, 
	{\it Run-and-tumble dynamics in a crowded environment : Persistent exclusion process for swimmers},
	\bibinfo{journal}{Physical Review E}, \textbf{\bibinfo{volume}{89}},
	\bibinfo{pages}{012706} (\bibinfo{year}{2014}).
	
	\bibitem[{\citenamefont{Kourbane-Houssene et~al.}}(2018)\citenamefont{Kourbane-Houssene, Erignoux, Bodineau and Tailleur}]{kourbanehoussene_2018}
	\bibinfo{author}{\bibfnamefont{R.}~ \bibnamefont{Kourbane-Houssene}}, 
	\bibinfo{author}{\bibfnamefont{C.}~\bibnamefont{Erignoux}},
	\bibinfo{author}{\bibfnamefont{T.}~\bibnamefont{Bodineau}} \bibnamefont{and}
	\bibinfo{author}{\bibfnamefont{J.}~\bibnamefont{Tailleur}}, 
	{\it Exact Hydrodynamic Description of Active Lattice Gases},
	\bibinfo{journal}{Physical Review Letters}, \textbf{\bibinfo{volume}{120}},
	\bibinfo{pages}{268003} (\bibinfo{year}{2018}).
	
	\bibitem[{\citenamefont{Whitelam et~al.}(2018)\citenamefont{Whitelam, Klymko and Mandal}}]{whitelam_2018}
	\bibinfo{author}{\bibfnamefont{S.}~ \bibnamefont{Whitelam}},
	\bibinfo{author}{\bibfnamefont{K.}~ \bibnamefont{Klymko}}, 
	\bibinfo{author}{\bibfnamefont{D.}~ \bibnamefont{Mandal}},
	{\it Phase separation and large deviations of lattice active matter},
	\bibinfo{journal}{Journal of Chemical Physics},
	\bibinfo{pages}{154902} (\bibinfo{year}{2018}).
	
	\bibitem{SI}
	Supplemental Material.	
	
\end{thebibliography}

\section{REFERENCES}
\begin{thebibliography}{100}
	\bibitem{Alefevre_jstatmech07}
	A. Lef\`{e}vre and G. Biroli, {\it Dynamics of interacting particle systems: stochastic process and field theory}, Journal of Statistical
	Mechanics: Theory and Experiment {\bf 2007}, P07024
	(2007)
	
	\bibitem{Alee_pre10}
	C. F. Lee, {\it Fluctuation-induced collective motion: A single-particle density analysis}, Physical Review E {\bf 81}, 031125 (2010).
	
	%
	%
	%
	%
	
	\bibitem{Achen_njp15} L. Chen, J. Toner, and C. F. Lee, {\it Critical phenomenon
		of the order-disorder transition in incompressible active fluids}, New Journal of Physics {\bf 17}, 042002 (2015).
	
	\bibitem{Abinder81} K. Binder, \textit{Finite Size Scaling Analysis of Ising Model Block Distribution Functions}, J. Phys. Zeitschrift f\"{u}r Physik B Condensed Matter \textbf{43}, 119-140 (1981)
	\bibitem{Arovere90} M. Rovere {\it et al.}, \textit{The Gas-Liquid Transition of the Two-Dimensional Lennard-Jones Fluid}, J. Phys. Condensed Matter \textbf{2}, 7009-7032 (1990)
	\bibitem{Asiebert2017} J. A. Siebert {\it et al.}, \textit{Critical Behaviour of Active Brownian Particles}, 
	Phys. Rev. E {\bf 98}, 030601 (2018)
	\bibitem{Alevis2014} D. Levis \& L. Berthier, \textit{Clustering and heterogeneous dynamics in a kinetic Monte Carlo model of self-propelled hard disks},  Phys. Rev. E \textbf{89}, 6 (2014)
	\bibitem{Awhitelam2018}  S. Whitelam {\it et al.}, \textit{Phase separation and large deviations of lattice active matter}, J. Chem. Phys. \textbf{148}, 154902 (2018)
	\bibitem{Aroman97} F. A. Roman {\it et al.}, \textit{Fluctuations in an equilibrium hard disk fluid: Explicit finite-size effects},  J. Chem. Phys. \textbf{107}, 4635 (1997)
	\bibitem{Arovere93} M. Rovere, P. Nielaba \& K. Binder, \textit{Simulation studies of gas-liquid transitions in two dimensions via a subsystem-block-density distribution analysis}, Z. Phys. \textbf{90}, 215-225 (1993)
	\bibitem{Aalexander94} F. A. Alexander \& D. A. Huse,  \textit{Dynamic Scaling and Decay of Correlations for Spinodal Decomposition at \(T_c\)}, Phys. Rev. B \textbf{50}, 2, (1994)
	\bibitem{Agonella2015} G. Gonnella {\it et al.},  \textit{Motility-induced phase separation and coarsening in active matter}, C. R. Physique \textbf{16} 316–331 (2015)
	\bibitem{Aredner2013} G. Redner {\it et al.},  \textit{Structure and dynamics of a phase-separating active colloidal fluid}, Physical Review Letters, \textbf{110}, 055701, (2013)
	\bibitem{Astenhammar2014} J. Stenhammar {\it et al.},  \textit{Phase behaviour of active Brownian particles: The role of dimensionality}, Soft Matter, \textbf{10}, 1489-1499, (2014)
\end{thebibliography}

\end{document}